\documentclass[12pt]{elsarticle}

\usepackage{lmodern}

\usepackage[version=4]{mhchem}

\usepackage[english]{babel}
\usepackage[utf8]{inputenc}
\usepackage[T1]{fontenc}

\usepackage{sidecap}

\usepackage[makeroom]{cancel}
\usepackage[numbers]{natbib}
\biboptions{sort&compress}
\usepackage{graphicx}
\usepackage{epstopdf}
\usepackage{rotating}
\usepackage{wrapfig}
\usepackage{float}
\usepackage{subfigure}
\usepackage{color}
\usepackage{hyperref}
\usepackage{datetime}
\usepackage{amsmath,amssymb,amsthm}
\usepackage{amsfonts}
\usepackage{mathtools}
\usepackage{cases}
\usepackage{array}
\usepackage[binary-units=true]{siunitx}
\usepackage{upgreek}

\usepackage[title]{appendix}

\usepackage[stable]{footmisc}

\usepackage{booktabs}
\usepackage{multirow}
\usepackage{multicol}
\usepackage{caption}

\usepackage{bm}

\usepackage{outlines}
\usepackage{dirtree}
\usepackage[dvipsnames]{xcolor}
\usepackage{framed}
\usepackage{color}
\usepackage{emptypage}
\usepackage{verbatim} 
\usepackage{textcomp}
\usepackage[section]{placeins}
\usepackage{pdfpages}
\usepackage{enumitem}
\setlist[enumerate]{itemsep=0mm}
\usepackage[flushleft]{threeparttable}
\usepackage{array,booktabs,makecell}
\usepackage{tcolorbox}
\usepackage{xargs}
\usepackage{tikz}
\usetikzlibrary{shapes}
\usepackage[textsize = tiny, color = blue!30]{todonotes}
\usepackage{marginnote}
\usepackage{lineno}
\usepackage[top=3.5cm, bottom=3.5cm, outer=3.5cm, inner=3.5cm, heightrounded, marginparwidth=3.cm, marginparsep=0.5cm]{geometry}

\usepackage{tcolorbox}
\usepackage{url}
\usepackage{epigraph} 
\usepackage[acronym]{glossaries}
\usepackage{soul}
\usepackage{wasysym}
\usepackage{physics}
\usepackage[textsize=tiny, color=blue!30]{todonotes}

\newtcolorbox{mybox}{colback=blue!5!white,colframe=blue!75!black}

\newcommand{\AF}[1]{{\color{blue} \textit{#1}}}

\newcommand{\AFbox}[1]{{ \begin{tcolorbox}[colback=blue!9!, colframe=blue!75!black] #1 \end{tcolorbox} } }

\newcommand{\AFpath}[1]{\AFbox{\path{#1} } } 

\newcommand{\AFmargin}[1]{{ \marginnote[]{\AF{#1}}[0.cm] }}

\newacronym{bb}{BB}{Blackbody}
\newacronym{bc}{BC}{Boundary Conditions}
\newacronym{bl}{BL}{Boundary Layer}
\newacronym{ccd}{CCD}{Charge-coupled device}
\newacronym{cfd}{CFD}{Computational Fluid Dynamics}
\newacronym{cmc}{CMC}{Ceramic Matrix Composite}
\newacronym{ceq}{CEQ}{Chemical EQuilibrium}
\newacronym{cneq}{CNEQ}{Chemical Non-EQuilibrium}
\newacronym{daq}{DAQ}{Data Acquisition Unit}
\newacronym{d4d}{D4D}{Design for Demise}
\newacronym{dsmc}{DSMC}{Direct-Simulation Monte Carlo}
\newacronym{eds}{EDS}{Energy-Dispersive X-Ray Spectroscopy}
\newacronym{eg}{EG}{Effective Gain}
\newacronym{em}{EM}{Electro Magnetic}
\newacronym{esa}{ESA}{European Space Agency}
\newacronym{ftbl}{FTBL}{Finite Thickness Boundary Layer}
\newacronym{fwhm}{FWHM}{Full Width at Half Maximum}
\newacronym{g2g}{G2G}{Gain To Gain}
\newacronym{geo}{GEO}{Geosynchronous Equatorial Orbit}
\newacronym{gsi}{GSI}{Gas-Surface Interaction}
\newacronym{gstp}{GSTP}{General Support Technology Program}
\newacronym{hf}{HF}{Heat Flux}
\newacronym{hsc}{HSC}{High-Speed Camera}
\newacronym{icp}{ICP}{Inductively Coupled Plasma}
\newacronym{ils}{ILS}{Instrument Line Shape}
\newacronym{ir3d}{IR3D}{InfraRed 3D}
\newacronym{irc}{IRC}{IR Camera}
\newacronym{ipm}{IPM}{Institute for Problems in Mechanics}
\newacronym{ir}{IR}{InfraRed}
\newacronym{irm}{IRM}{Instrument Response Model}
\newacronym{irs}{IRS}{Institut f\"ur Raumfahrtsysteme}
\newacronym{leo}{LEO}{Low-Earth Orbit}
\newacronym{lif}{LIF}{Laser Induced Fluorescence}
\newacronym{lhts}{LHTS}{Local Heat Transfer Simulation}
\newacronym{los}{LOS}{Line Of Sight}
\newacronym{lte}{LTE}{Local Thermodynamic Equilibrium}
\newacronym{lut}{LUT}{Look-Up Table}
\newacronym{mhd}{MHD}{Magneto-Hydro-Dynamics}
\newacronym{nasa}{NASA}{National Aeronautics and Space Administration}
\newacronym{ndp}{NDPs}{Non-Dimensional Parameters}
\newacronym{nir}{NIR}{Near Infra Red}
\newacronym{nist}{NIST}{National Institute of Standards and Technology}
\newacronym{ode}{ODE}{Ordinary Differential Equation}
\newacronym{oes}{OES}{Optical Emission Spectroscopy}
\newacronym{oc}{OC}{Optical Calibration}
\newacronym{op}{OP}{Optical Path}
\newacronym{plc}{PLC}{Programmable-Logic Control}
\newacronym{pwt}{PWT}{Plasma Wind Tunnel}
\newacronym{ptb}{PTB}{Plasmatron Test Bench}
\newacronym{rh}{RH}{Relative Humidity}
\newacronym{rms}{RMS}{Root Mean Square}
\newacronym{rss}{RSS}{Residual Sum of Squares}
\newacronym{rhs}{RHS}{Right-Hand Side}
\newacronym{sd}{SD}{Space Debris}
\newacronym{sem}{SEM}{Scanning Electron Microscopy}
\newacronym{snr}{SNR}{Signal to Noise Ratio}
\newacronym{smf}{SMF}{Spatial Magnification Factor}
\newacronym{srf}{SRF}{Spatial Resolution Function}
\newacronym{surf}{SURF}{Electrochemical and Surface Engineering research group}
\newacronym{talif}{TALIF}{Two-photon Absorption Laser Induced Fluorescence}
\newacronym{tc}{TC}{Thermocouple}
\newacronym{tcp}{TCP}{Two-Color Pyrometry}
\newacronym{tps}{TPS}{Thermal Protection System}
\newacronym{uv}{UV}{Ultra Violet}
\newacronym{vki}{VKI}{von Karman Institute for Fluid Dynamics}
\newacronym{vub}{VUB}{Vrije Universiteit Brussel}

\newacronym{hs30}{HS30}{Hemisphere probe 30 mm diameter}
\newacronym{hs50}{HS50}{Hemisphere probe 50 mm diameter}
\newacronym{st50}{ST50}{Standard probe 50 mm diameter}

\newacronym{neqair}{NEQAIR}{Non-EQuilibrium Air Radiation code}
\newacronym{coolfluid}{CF-ICP}{CoolFLUID ICP-code}
\newacronym{cficp}{CF-ICP}{CoolFLUID ICP-code}
\newacronym{neboula}{NEBOULA}{Non-Equilibrium Boundary Layer solver}
\newacronym{cerbere}{CERBERE}{Catalicity and Enthalpy Reuilding for a Reference Probe}

\newcommand{\cficp}[0]{{\texttt{CF-ICP}}}
\newcommand{\cficppr}[0]{{\texttt{CF-ICP-PR}}}

\newcommand{\suprm}[1]{{^{\mathrm{#1} } }} 
\newcommand{\subrm}[1]{{_{\mathrm{#1} } }} 
\newcommand{\subsuprm}[2]{{ _{\mathrm{#1}}^{\mathrm{#2} } }}

\DeclareSIUnit[per-mode=symbol,per-symbol=p]{\us}{\micro\second}
\DeclareSIUnit[per-mode=symbol,per-symbol=p]{\ms}{\milli\second}

\DeclareSIUnit[per-mode=symbol,per-symbol=p]{\kW}{\kilo\watt}
\DeclareSIUnit[per-mode=symbol,per-symbol=p]{\MW}{\mega\watt}

\DeclareSIUnit[per-mode=symbol,per-symbol=p]{\mbar}{\milli\bar}

\DeclareSIUnit[per-mode=symbol,per-symbol=p]{\nm}{\nano\meter}
\DeclareSIUnit[per-mode=symbol,per-symbol=p]{\um}{\micro\meter}
\DeclareSIUnit[per-mode=symbol,per-symbol=p]{\cm}{\centi\meter}
\DeclareSIUnit[per-mode=symbol,per-symbol=p]{\angstrom}{\mathring{A}}

\newcommand{\cm}[0]{{~\si{\centi\metre}}} 
\newcommand{\mm}[0]{{~\si{\milli\metre}}} 
\newcommand{\um}[0]{{~\si{\micro\metre}}}  
\newcommand{\nm}[0]{{~\si{\nano\metre}}} 

\DeclareSIUnit\torr{Torr}
\DeclareSIUnit\psiunit{psi}

\newcommand*{\parderconst}[3]{\left( \frac{\partial#1}{\partial#2} \right)_{#3} }

\newcommand{\boltzmannconstant}{k_{\mathrm{B}}}

\newcommand{\reynolds}{Re}

\newcommand{\DTwat}{\Delta T_{\ce{H2O}} }
\newcommand{\qw}{\dot{q}_{\textrm{w}} }

\newcommand{\lambdap}[0]{\lambda_{p}}
\newcommand{\Lpix}[1][]{\bar{L}_{\lambdap}^{\textrm{#1}} }
\newcommand{\Uhatpix}[1][]{\hat{U}_{p}^{\textrm{#1}}}

\newcommand{\Phitildezero}[1][]{\tilde{\Phi}_{0 } }

\newcommand{\boltzmanncoeff}{\ln\left( \frac{n_{\textrm{u}, i}}{g_u}\right)}
\newcommand{\einsteincoeff}{\mathcal{A}_{ul}}

\newcommand{\eps}{\varepsilon}

\newcommand{\gaussian}{\mathcal{G}}
\newcommand{\lorentzian}{\mathcal{L}}
\newcommand{\voigt}{\mathcal{V}}

\newcommand{\specnot}[5]{ \textrm{#1} \; {}^{#2}{\textrm{#3}}_{#4}^{#5}}

\newcommand{\Hbeta}{\ce{H}_{\upbeta}}

\newcommand{\Tedge}{T\subrm{e}}
\newcommand{\pedge}{p\subrm{e}}

\newcommand{\hedge}{h\subrm{e}}
\newcommand{\betaedge}{\beta\subrm{e}}
\newcommand{\gammaprime}[1][]{\gamma'}
\newcommand{\gammaref}[1][]{\gamma\subsuprm{ref}{#1}}
\newcommand{\gammaw}[1][]{\gamma\subsuprm{w}{#1}}

\newcommand{\Ts}{T\subrm{s}}
\newcommand{\us}{u\subrm{s}}

\newcommand{\hs}{h\subrm{s}}
\newcommand{\rhos}{\rho\subrm{s}}
\newcommand{\mus}{\mu\subrm{s}}
\newcommand{\deltastar}{\delta^{*}}

\newcommand{\Twall}{T_{\mathrm{w}}} 		

\newcommand{\pc}{p_{\mathrm{c}}} 			

\newcommand{\pdyn}{p_{\mathrm{dyn}}} 		

\newcommand{\qcw}{\dot{q}_{\mathrm{cw}}} 	
\newcommand{\mdot}{\dot{m}_{\mathrm{gas}}} 	
\newcommand{\mdotwat}{\dot{m}_{\ce{H2O}}} 	
\newcommand{\Pel}{P_{\mathrm{el}}}	 		
\newcommand{\Pelsim}{P_{\mathrm{el}}^{\textrm{sim}}}	 		
\newcommand{\Pelmeas}{P_{\mathrm{el}}^{\textrm{meas}}}	 		

\newcommand{\Ntwopfirstneg}{\ce{N2+(B^2\Sigma^+_u \rightarrow X^2\Sigma_g)}} 
\newcommand{\Ntwosecondpos}{\ce{N2(C^3\Sigma_u \rightarrow B^3\Sigma_g)}} 
\newcommand{\Ntwofirstpos}{\ce{N2(B^3\Sigma_g \rightarrow A^3\Sigma_g^+)}} 
\newcommand{\CNvio}{\ce{CN(B^2\Sigma^+ \rightarrow X^2\Sigma^+)}}

\begin{document}

\begin{frontmatter}

\title{Characterization of the VKI Plasmatron subsonic ICP jet combining optical emission spectroscopy, intrusive measurements, and CFD simulations}
\date{}
\author[inst1,inst3]{Andrea Fagnani\corref{cor1}\fnref{fn1}}
\ead{andrea.fagnani@vki.ac.be}
\cortext[cor1]{Corresponding author}
\fntext[fn1]{Current affliation: NASA Postdoctoral Fellow at NASA Ames Research Center, Moffett Field, CA 95034, USA}

\author[inst1]{Bernd Helber}
\ead{bernd.helber@vki.ac.be}

\author[inst1]{Damien Le Quang}
\ead{damien.lequang@vki.ac.be}

\author[inst1,inst4]{Alessandro Turchi}
\ead{alessandro.turchi@asi.it}

\author[inst1]{Jimmy Freitas Monteiro}
\ead{jlfmonteiro89@hotmail.com}

\author[inst3]{Annick Hubin}
\ead{annick.hubin@vub.be}

\author[inst1]{Olivier Chazot}
\ead{olivier.chazot@vki.ac.be}

\affiliation[inst1]{organization={Aeronautics and Aerospace Department, von Karman Institute for Fluid Dynamics},
            addressline={Chaussée de Waterloo 72}, 
            city={Rhode-st-Genèse},
            postcode={1640}, 
            country={Belgium}}

\affiliation[inst3]{organization={Materials and Chemistry Department, Vrije Universiteit Brussel},
	addressline={Plainlaan~2}, 
	city={Brussel},
	postcode={1150}, 
	country={Belgium}}
\affiliation[inst4]{organization={Italian Space Agency, Science and Research Directorate},
         	addressline={via del Politecnico}, 
         	city={Rome},
         	postcode={00133}, 
         	country={Italy}}

\begin{abstract}
	This paper addresses the characterization of the subsonic flow in the 1.2~MW Inductively Coupled Plasma (ICP) wind tunnel at the von Karman Institute for Fluid Dynamics (VKI), targeting chamber pressures of 50 and 100~mbar, and input electric powers between 150 and 300~kW. 
	Ultraviolet to near-infrared optical emission spectroscopy measurements of the free-jet flow are carried out with an updated experimental set-up, calibration procedure, and data processing, providing high-quality absolute spatially-resolved emission spectra. 
	Emission measurements agree with thermochemical equilibrium predictions within a range of conditions, allowing to extract experimental maps of cold-wall heat flux and dynamic pressure against the inferred free-jet enthalpy.
	A detailed comparison with the characterization methodology traditionally employed is presented, highlighting the need for an improved modeling strategy.
	Using the measured free-jet temperature and dynamic pressure only, a forward procedure for the computation of the stagnation line flow is proposed. 
	The latter agrees with intrusive heat flux measurements through a range of test conditions, and for values of the recombination coefficient of the reference copper probe commonly found in the literature.
	Results demonstrate that a consistent framework between numerical simulations and experimental data can be achieved, defining an improved framework for the characterization of the subsonic ICP jet.

\end{abstract}


\end{frontmatter}



\section{Introduction}
\label{sec:OES:intro}
\graphicspath{{figures/}}

The 1.2 MW Plasmatron facility at the von Karman Institute for Fluid Dynamics (VKI) is an Inductively Coupled Plasma (ICP) Wind Tunnel (PWT) capable of providing experimental duplication of stagnation line boundary layer flows experienced by a spacecraft during planetary entry \cite{Bottin2000}. Test conditions are suitable to study gas-surface interaction phenomena, relevant to the response of candidate materials to atmospheric entry flows. These include thermal protection materials \cite{Panerai2012a, Helber2016b}, used to shield spacecraft against shock layer heating, and space debris components \cite{josephelrassi2023a}, whose aerothermal demise shall be ensured according to recently proposed post-mission disposal strategies.

The quality of the experimental material response analysis in PWTs is largely dependent on the accurate characterization of the flow. This is important for scaling the ground test to an equivalent flight condition \cite{Barbante2006, Turchi2021}, as well as for the precise numerical reproduction of the experiment and its comparison to in situ measurements of the material response \cite{Fagnani2021, Fagnani2023a}. This step is necessary to validate and improve computational models, subsequently employed to predict the material behavior for a real case scenario.

In this context, conventional procedures for the plasma flow characterization are often based on inverse heat-transfer methods, which estimate the gas enthalpy and velocity from intrusive measurements of heat flux and Pitot pressure \cite{Degrez2001, Park2006, Viladegut2020}. These typically employ semi-empirical correlations, or numerical simulation procedures, under the assumption of Local Thermodynamic Equilibrium (LTE) in the free-jet flow. Such methods are inherently affected by uncertainties propagating from the measured quantities, and systematic errors in the choice of the required model parameters \cite{Turchi2017a, Sanson2018}. 
Additionally, a consistent experimental validation of such inverse approach is often missing.

This work details the recent advancements in the flow characterization capabilities achieved in the VKI Plasmatron.  
The experimental data leverage ultraviolet to near-infrared Optical Emission Spectroscopy (OES) measurements of the subsonic free-jet plasma flow, expanding the work initially presented in Ref. \cite{Fagnani2020a}. The updated experimental set-up, calibration procedure, and data processing provide high quality absolute spatially-resolved emission spectra. Radial temperature profiles are extracted from atomic lines and spectral fits, and radial electron density profiles are measured from the Stark broadening of the hydrogen Balmer-beta line. 
Measurements support the LTE assumption within a range of conditions, allowing an independent estimate of the flow enthalpy. 

Temperature profiles are compared against 2D axisymmetric CFD simulations of the flow field within the test chamber using an in-house magnetohydrodynamics code. This gives an estimate of the numerical power efficiency required, and highlights both relevant consistencies and discrepancies with the simulation results.
We provide experimental maps of cold-wall heat flux and jet dynamic pressure for different probe radii, as a function of the inferred free-jet (FJ) flow enthalpy.
Data are compared against the traditional inverse rebuilding procedure employed in the facility, highlighting the need for improvements.

Based on the measured free-jet temperature and dynamic pressure, a forward procedure for the computation of the stagnation line flow is proposed. The method employs a quasi one-dimensional formulation of the Navier-Stokes equations, instead of the commonly used boundary layer approximation, thus requiring fewer parameters than what was previously needed.
The new procedure demonstrates compatible predictions with the measured cold-wall heat flux through a range of the tested conditions, and for values of the recombination coefficient of the copper probe found in the literature.

Our results provide a step towards an improved characterization of the subsonic ICP  flow, showing that a consistent framework between numerical simulations and experimental data can be achieved within a range of the facility test envelope.

\section{Enthalpy characterization methods for PWTs and their limitations}
\label{sec:h_methods}

While a complete duplication of re-entry flight conditions is not feasible around a scaled model in a ground facility, a local duplication can be achieved around the stagnation-point region. 
The Boundary Layer (BL) equations represent the main modeling framework to derive the similarity parameters between a real flight condition and an experimental simulation in a ground test facility \cite{Fay1958, Goulard1958, Rose1958}. 
In particular, the Local Heat Transfer Simulation (LHTS) methodology, discussed originally by Kolesnikov~\cite{Kolesnikov1993, Kolesnikov1999, Kolesnikov2000}, specifically addresses the application to ICP wind tunnels. The latter leverages the parabolic nature of the BL problem to identify sufficient parameters to locally duplicate a flight condition in a ground test framework. Under the assumption of steady-state conditions and LTE at the BL edge, these are represented by the gas enthalpy $\hedge$ and pressure $\pedge$ at this point, the inviscid radial velocity gradient at the wall $\hat{\beta}_\subrm{w}$ \cite{Turchi2021}, as well as the conditions at the material's surface.
The extent of the LHTS was subsequently analyzed for arbitrary catalytic surfaces by \citet{Barbante2006}, and additionally extended by \citet{Turchi2021} to the case of materials undergoing ablation.

In particular, the accurate determination of $\hedge$ in PWTs represents a challenging task, and several approaches have been reported in the literature \cite{Park2006}. Among these, the inverse heat transfer method is based on intrusive measurements of heat flux and pressure at the stagnation point on a reference probe, and likely represents the most commonly adopted among different facilities. 

Relevant to this category is the semi-empirical method initially published by Zoby and Sullivan \cite{Zoby1968} and Pope \cite{Pope1968}, which is now included as a standard technique in the ASTM E637 \cite{ASTM-E637}.
The VKI enthalpy rebuilding procedure, instead, is based on a numerical solution of the BL equations through an iterative approach. The reader is addressed to the works of \citet{Degrez2001}, \citet{Barbante2006}, \citet{Turchi2017} and \citet{Viladegut2020} for a comprehensive description of this methodology.

In both cases, the functional form of the wall heat flux, $\qw$, can be written as
	\begin{equation}
		\qw = \qw(\Tedge, \pedge, \delta\subrm{e},  u\subrm{e},\beta_e,  \beta_e', \Twall, \gamma\subrm{w}, k_r),
	\end{equation}
highlighting the dependence on the BL edge quantities, in terms of temperature $ T\subrm{e} $, pressure $ p\subrm{e} $, BL thickness $\delta\subrm{e}$, axial velocity $ u\subrm{e} $, radial velocity gradient $ \beta_e $ and its axial derivative $ \beta_e' $, as well as on the surface temperature $ \Twall $,  wall catalytic efficiency $ \gamma\subrm{w}$, and the gas chemical rates $k_r$. 
Starting from the measured wall heat flux and stagnation pressure, the method estimates the BL edge temperature, and, hence, the flow enthalpy, under the assumption of LTE at the BL edge.

In this regard, the rebuilt value of $\hedge$ is known to be particularly sensitive to the modeling assumptions on $ \gamma\subrm{w} $ and $ k_r $, and further degraded by experimental uncertainties on $\qw $ and $ u\subrm{e} $ \cite{Panerai2012, Turchi2017}.
Copper is traditionally used to perform calorimetric measurement in high-enthalpy wind tunnels, as it provides both a high catalytic efficiency and a high thermal conductivity. 
The former often leads to the assumption of a fully catalytic surface, within a conservative approach employed by several authors in the context of TPS ground testing \cite{Degrez2001, Nawaz2013}. 
A more precise characterization of $ \gammaw $ was addressed in detailed studies on the recombination efficiency of copper and its oxides in $ \ce{O2} $, $ \ce{N2} $, and air atmospheres in the literature \cite{Linnett1956, Greaves1958, Greaves1959, Young1961, Rosner1963, Dickens1964, Hartunian1965, Winkler1966, Pope1968a, Melin1971, Rosner1977, Cauquot1998, Chazot2008, Nawaz2013, Park2013, Driver2015, Fletcher2017, Viladegut2017, Viladegut2020, Yang2021}.
\citet{Nawaz2013}, in particular, have shown that copper is rapidly oxidized to $\ce{CuO}$ when exposed to a dissociated oxygen atmosphere, for which available data of recombination coefficients are mostly found in the range between $ 10^{-2} $ and $ 10^{-1} $. 

For the VKI Plasmatron facility, Panerai \cite{Panerai2012} suggested the use of $\gammaw=0.1 $ for $\SI{12}{\mbar} < \pc \leq  \SI{50}{\mbar}$ and $\gammaw = 0.01 $ for $\SI{50}{\mbar} < \pc \leq \SI{100}{\mbar}$ when using the ESA Standard Probe geometry employing a copper calorimeter. These values were subsequently adopted in several studies \cite{Helber2016, Sakraker2016}.
More recently, \citet{Viladegut2020} estimated the value of $ \gammaw $ through comparative experiments in the VKI Plasmatron, providing a range between $6.36 \times 10^{-3}$ and $ 8.82 \times 10^{-2}$ for pressures between $\SI{200}{\mbar}$ to $\SI{15}{\mbar}$.

However, achieving accurate measurements of $\gammaw$ in PWT experiments is challenging, due to its dependence on the specific chemical state of the catalyst \cite{Cauquot1998}, partial pressure of the impinging gas \cite{Viladegut2022}, incomplete energy accommodation at the surface \cite{Rosner1977}, and the coupled influence of diffusion-reaction effects \cite{Rosner1963}.
Moreover, estimates of $\gammaw$ from PWT experiments are usually implicitly related to the free-jet characterization methodology employed and, hence, deemed less accurate.
Additionally, the inferred flow enthalpy is often not cross-validated with different techniques, and the LTE assumption at the BL edge remains questionable for a range of conditions in which the ICP torch can be operated.

It is the purpose of this work to specifically address the problem, providing relevant experimental data from OES for comparison to the existing method, and to develop a numerical-experimental methodology that is independent of the reference catalytic coefficient.

\section{Experimental}
\label{sec:OES:setup}

\subsection{The VKI Plasmatron facility}
The VKI Plasmatron features a $ \SI{160}{\milli\meter} $ diameter ICP torch, powered by a $ \SI{400}{\kilo \hertz} $, $ \SI{1.2}{\mega \watt} $, $ \SI{2}{\kilo \volt} $ electric generator, and connected to a $ \SI{1.4}{\meter} $ diameter, $ \SI{2.4}{\meter} $ long test chamber.  An extensive description of the facility and its performance was given by Bottin~et~al.~\cite{Bottin2000}.
Fig.~\ref{fig:figure3_1} shows a schematic section of the ICP torch, test chamber, and instrumentation set-up described in the following sections.
The torch is made up of a quartz tube, surrounded by a six-turn flat coil inductor, and supplied by a gas injection system. The electric power to the coil, $ \Pel $, is monitored by a voltage-current probe, while a calibrated flow meter (F-203AV, Bronkhorst High-Tech B.V, NL) controls the mass flow rate, $ \mdot $, of the test gas supplied to the torch with a $ 0.5\% $ accuracy. For all the conditions reported in this study, $\mdot$ was fixed at $\SI{16}{\gram \per \second}$. The test gas, either compressed atmospheric air or synthetic mixtures, is heated by electromagnetic induction to provide a chemically pure plasma flow. Pressure in the test chamber, $ \pc $, is measured by an absolute pressure transducer (Membranovac DM 12, Leybold GmbH, DE) to an accuracy of 1\%.

\subsection{Intrusive measurements of heat flux and dynamic pressure}

Three movable holding arms can be swung into the plasma flow interchangeably by a pneumatic mechanism, holding copper-cooled probes with 50 and 30~mm diameter hemispherical heads, named HS50 and HS30, respectively, in the context of this work.
A $ \SI{14}{\mm} $ diameter copper water-cooled calorimeter was used to measure the stagnation point Heat Flux (HF), $ \qcw $, as
\begin{equation}
	\qcw = \frac{\mdotwat \cdot c_p \cdot \DTwat}{A},
\end{equation}
where  $ c_p $ is specific heat of water, and $ A $ is the surface area.
The water mass flow rate, $ \mdotwat $, is measured by two liquid dosing units (M14-RGD-33-0-S, Bronkhorst, NL), fed through a demineralized closed-loop system, while type-E thermocouples measure the inlet-outlet temperature difference, $ \DTwat $.
The uncertainty on $\qcw$ is estimated to $ \pm10\% $ at $ \qcw = \SI{500}{\kW / \meter\squared} $, decreasing to about 6\% above $ \qcw = \SI{2000}{\kW / \meter\squared} $. The accuracy of $\qcw$ measurements was additionally verified on a calibration bench with the help of an electric heater and a thermal balance approach.

Stagnation pressure was measured through a cooled HS50 Pitot probe. Considering that the Bernoulli equation is a valid first-order approximation for low Mach number plasma flows~\cite{Magin2001}, a low-Reynolds number correction is accounted for with the pressure coefficient following the analytical derivation by Homann~\cite{Homann1952}
\begin{equation}
	\begin{cases}
		C_p = \frac{\pdyn}{1/2 \rhos \us^2} = 1 + \frac{6}{\reynolds\subrm{s} + 0.455 \sqrt{\reynolds\subrm{s}}} \\
		\reynolds\subrm{s} = \rhos \us R / \mus,
	\end{cases}
	\label{eq:pitot}
\end{equation}  
$\pdyn$ being the dynamic pressure, measured as the difference between the stagnation and chamber pressures by a Validyne differential pressure transducer  (DP45, Validyne Engineering Corp., USA), while $\rhos$, $\us$ and $\mus$ are the flow density, axial velocity and viscosity, respectively, and $R$ is the probe radius.
Due to the lack of a uniform upstream flow within the jet, the free-jet quantities, indicated with the subscript "$\textrm{s}$" in the previous equation, are defined as the point in the free-jet flow at the location where the probe would be injected, i.e.,  385~mm from the torch exit. 
The uncertainty on  $ \pdyn $ is estimated to $ \pm \SI{5}{\pascal} $, and mostly related to the oscillations experienced during measurements.
A 30-channel data acquisition unit (MX100, Yokagawa, JP), allows recording of thermocouple signals and output voltage from different instruments through a computer interface.

\subsection{Emission spectroscopy set-up}

\begin{figure}
	\centering
	{\includegraphics[trim = {0cm, 2cm, 13cm, 2cm}, clip, width=.95\textwidth]{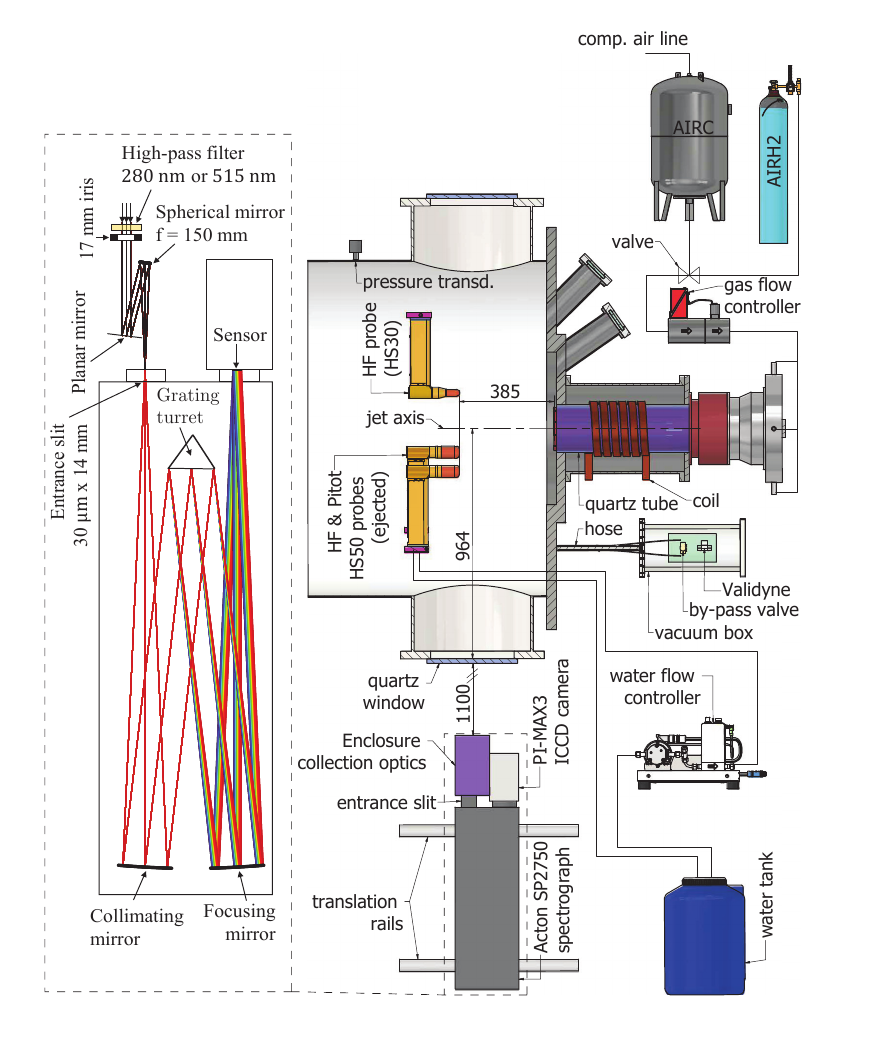}} \\
	
	\caption[Schematic of the OES set-up for flow characterization.]{Schematic of the experimental set-up designed for this work, showing the VKI Plasmatron chamber and the instrumentation employed for the free-jet flow characterization (some components and lengths are not to scale for illustration purposes), along with a close-up view of the emission spectroscopic system.}
	\label{fig:figure3_1}
\end{figure}

Figure~\ref{fig:figure3_1} also illustrates the details of the OES set-up.
A $ \SI{750}{\mm} $ focal-length Czerny-Turner spectrograph (Acton SP-2750, Princeton Instruments, USA), coupled to a $1024\times1024$ pixel intensified CCD detector (PI-MAX3, Teledyne Princeton Instruments, USA), was used to record spatially resolved spectra of the free-jet plasma flow, allowing the light to be chromatically resolved along each row, and spatially resolved along each column of the detector. A 150 grooves/mm grating was employed for broad-range ultraviolet (UV) to near-infrared (NIR) spectra, while a $ \SI{1200}{grooves / \mm} $ grating allowed higher resolution measurements of the $ \Hbeta $ lineshape. An entrance slit width of $ \SI{30}{\micro\meter} $ was selected to optimize the spectral resolution, while the intensifier gain and gate time were adjusted to provide a high signal to noise ratio for each measurement.

A major update of the OES set-up included a new optical train, featuring a 150~mm focal length spherical mirror, and a planar fold mirror, ensuring chromatic aberration free measurements over the measured wavelength range.
The optics stood at 2.4~m from the plasma jet axis, providing an optical magnification of about $ 15.1 $.
A $ \SI{17}{\mm} $ diameter iris avoided stray light, slightly overfilling the optical throughput of the spectrograph. High-pass filters with cut-on wavelengths at $ \SI{280}{\nm} $ and $ \SI{515}{\nm} $ were selected to avoid contamination by second order dispersion, according to the measurement range.
The optical bench was mounted on translation rails to perform measurement at different axial positions along the jet axis, i.e., at 195 and $ \SI{385}{\mm} $ from the torch exit, the latter being the location where probes were injected in the flow.

The spatial calibration provided a magnification factor of $\SI{0.194}{\mm / pixel}$ over a domain size of $ \SI{198.4}{\mm} $, while the spatial resolution function was fit with a Gaussian profile of $\SI{3.071}{\mm}$ at Full-Width Half Maximum (FWHM). Simulations demonstrated negligible effect of the spatial smearing of the measured radiance profiles for the conditions investigated in this work.
Wavelength calibration against low-pressure $ \ce{Hg} $ and $ \ce{Ar}\textsf{-}\ce{Ne} $ lamps gave a resolution of $ \SI{0.112}{\nm/pixel} $ and $ \SI{0.013}{\nm/pixel} $ at $\SI{500}{\nm}$ for the $ \SI{150}{grooves/\mm} $ and $ \SI{1200}{grooves/\mm} $ gratings, respectively.
The square root of a Voigt profile \cite{Brandis2016, Cruden2014a} demonstrated superior fitting of the Instrument Line Shape (ILS) compared to standard Gaussian or Voigt functions, and the fit parameters are listed in Table~\ref{tab:OES_ILS_FWHM}.

\begin{table}[]
	\centering
	\caption{FWHM of the fitted ILSs ($\Delta \lambda_{\sqrt{\mathcal{V}}}$) and their Gaussian ($\Delta \lambda_{\mathcal{G}}$) and Lorentzian ($\Delta \lambda_{\mathcal{L}}$) components.}
		\begin{tabular}{lllll}
			\hline
			\vspace{0.5pt}
			 grating & $ \lambda_0 $ & $ \Delta \lambda_{\mathcal{G}}  $ & $ \Delta \lambda_{\mathcal{L}}  $  & $ \Delta \lambda_{\sqrt{\mathcal{V}}} $ \\ 
			 grooves/mm & nm & nm & nm & nm \\ \hline
			 150 & 435.8 & $  0.3506 $ & $ 0.0164 $ & $ 0.5104  $ \\
			 150 & 763.5 & $ 0.2885  $ & $ 0.0179  $ & $ 0.4239  $ \\ 
			 1200 & 435.8 & $ 0.0334  $ & $ 0.0015  $ & $ 0.0485   $\\ 
			 \hline
		\end{tabular}
	\label{tab:OES_ILS_FWHM}
\end{table}

Intensity calibration was performed against a tungsten ribbon lamp (Osram WI 17/G) traceable to NIST. Assuming linearity of the dark subtracted digitized intensity counts $\Uhatpix$ with respect to both the incident radiance and exposure time, the measured radiance during experiment was found as
\begin{equation}
	\Lpix[m] =  \frac{\Lpix[c]}{\Uhatpix[c] / \Delta t\suprm{c}} \times \frac{\Uhatpix[m]}{\Delta t\suprm{m}},
	\label{eq:OES_meas}
\end{equation}
where $ \Lpix[c] $ is the known spectral radiance of the calibration source at the wavelength $ \lambdap $, $\Delta t$ indicates the gate time, while the subscripts "c" and "m" refer to the calibration and measurement quantities, respectively. Attention was paid to avoid saturation of the intensifier occurring at high gains during calibration, and the accuracy of the radiance calibration was additionally validated against a high-temperature blackbody source.

\subsection{OES data processing}
\label{sec:OES:techniques}

After calibration, each pixel of the acquired frame yields the radiance, $ L(\lambda, y) $, integrated along a line-of-sight crossing the plasma jet, resolved in the spectral and radial dimensions at the measurement location along the axis.
Under the assumptions of axisymmetry and optically thin radiance, the local spectral emission intensity, $ \eps(\lambda, r) $ as a function of the radial coordinate $r$ can be obtained from $ L(\lambda, y) $ through the inverse Abel transform \cite{Kunze2009}. The latter is solved numerically using the PyAbel Python package \cite{Hickstein2019}, after spatial filtering and symmetrization. 

The temperature corresponding to excited atomic states is obtained from the optically thin emission intensity \cite{Kunze2009}
\begin{equation}
	\eps_{ul} = \frac{E_u - E_l}{4 \uppi} \einsteincoeff n_{i, u},
	\label{eq:OES_line_emission}
\end{equation}
where $ E_u $ and $ E_l $ are the upper and lower state energies, $ \einsteincoeff $ is the Einstein coefficient of spontaneous emission for the transition $ u \rightarrow l $, and $ n_{i, u} $ is the number density of the state $ u $ of species $ i $. For a LTE plasma, a Boltzmann population of the energy levels exists
\begin{equation}
	n_{i, u} = n_i(p, T) g_{i, u} \frac{e^{-E_{i, u} / (\boltzmannconstant T)}}{Q_{i, \textrm{int}}(T)},
	\label{eq:OES_boltzmann_population}
\end{equation}
with $n_i$ being the number density of the species, $g_{i, u}$ the degeneracy of the level,  $Q_{i, \textrm{int}}$ the internal partition function and $\boltzmannconstant$ the Boltzmann constant.
The NIST database \cite{NISTASD} provided the necessary spectroscopic constants, reported in table~\ref{tab:APP_OES_atomic_transitions}, and partition functions, while the Chemical EQuilibrium (CEQ) composition of the gas, $  n_i(p, T) $, was computed with the $\textrm{Mutation}^{++}$ library \cite{Scoggins2020}.
The low subsonic flow characteristics of VKI Plasmatron subsonic ICP torch allow to assume a uniform pressure within in the chamber, i.e., $ p \approxeq \pc $. Then, $ \eps_{ul} \approxeq \eps_{ul}(\pc, T) $ could be readily inverted for $T$ from the measured $\eps_{ul}$. For closely spaced multiplet transitions, the total emission intensity is $ \eps\subrm{tot} = \sum_{u, l \in U, L}  \eps_{ul}$, and a linear baseline subtraction was considered to isolate the atomic emission intensity from the background molecular or continuum radiation.

Alternatively, dividing eq.~\ref{eq:OES_boltzmann_population} by $ g_u $ and taking the logarithm yields
\begin{equation}
	\boltzmanncoeff = \ln\left(\frac{n_i (p, T)}{Q_{\textrm{int}, i}(T)} \right)  - \frac{E_u}{ \boltzmannconstant T},
	\label{eq:boltzmann_plot}
\end{equation}
where the $n_{i, u}$ is obtained from the measured emission intensity upon inversion of eq.~\ref{eq:OES_line_emission}.
Hence, for a Boltzmann population of the excited states, the coefficient $ \ln ( 	n_{\textrm{u}, i } / {g_u} ) $, is linear with respect to the upper energy $ E_u $, the angular coefficient being inversely proportional to the electronic temperature. 

A LTE temperature corresponding to the excited state transitions of molecular systems was additionally estimated by fitting the experimental spectra with the ones simulated by NEQAIR v15.0 \cite{whiting1996,Cruden2014}, employing a single temperature and a Boltzmann population model. The radiation model, developed at NASA Ames Reseach Center, is a line-by-line code that computes spontaneous emission, absorption and stimulated emission due to transitions between various energy states of excited atomic and molecular species along a line-of-sight through a non-uniform gas mixture. 
A database of synthetic spectra was generated at the pressures of interest with a temperature interval of $\SI{100}{\kelvin}$ and convolved with the experimental ILS. 
Simulations using NEQAIR also confirmed negligible self-absorption at the conditions considered in this work.

Finally, the electron number density was measured from the Stark broadening of the Balmer beta line of the hydrogen atom at $ \SI{486.1}{\nm} $, indicated with $\Hbeta$. The latter offers a strong and linear Stark effect, with a low sensitivity to ion dynamics perturbations, and negligible self absorption, allowing measurements of electron densities as low as $ \SI{1e14}{\per \cm \cubed} $ \cite{Gigosos1996, Yubero2005}. To this purpose, we used a mixture of synthetic air with 2\% $\ce{H2} $ ($ 77.91\%\;\ce{N2} $, $ 20.09\%\;\ce{O2} $, $ 2.00\%\; \ce{H2} $ in volume), indicated as AIRH2, to provide enough line strength of the $ \Hbeta $ line. 
A correlation for the Stark broadening of the $ \Hbeta $ line as a function of the electron number density was built from the electrodynamics simulations of \citet{Gigosos1996}.
Doppler, resonance, natural and van deer Waals broadening correlations are taken from Laux~et~al.~\cite{Laux2003}, based on the analytical expressions from Griem~\cite{Griem2005}.

Considering that each line $ j $ of the multiplet is a Voigt profile centered at the transition wavelength $ \lambda_{0, j} $ as $ \voigt_j(\lambda) = \voigt(\lambda -\lambda_{0, j}, \Lambda_{\lorentzian}, \Lambda_{\gaussian})  $, the simulated transition, $ \mathcal{S}^{*}(\lambda) $, will be given by the convolution of their sum with the experimental ILS.
The electron number density was found by matching the computed FWHM to the experimental value. 
In so doing, the measured chamber pressure and the LTE temperature inferred from the atomic lines was used to compute the other broadening mechanisms.

\section{Numerical simulation tools}
\subsection{VKI CF-ICP code}

The subsonic flow in the Plasmatron chamber was numerically simulated using a two-dimensional magnetohydrodynamic solver, referred to as VKI CF-ICP code in the following, which couples the Maxwell induction equations with the Navier-Stokes equations under the assumptions of LTE and axisymmetric steady flow \cite{Degrez2004}. The code is integrated into the Computational Object-Oriented Library for Fluid Dynamics (COOLFluiD) \cite{Lani2013} and relies on the Mutation++ library \cite{Scoggins2020} to compute the thermodynamic and transport properties of an eleven-species air mixture ($ \ce{O2} $, $ \ce{N2} $, $ \ce{O2+} $, $ \ce{N2+} $, $ \ce{NO} $, $ \ce{NO+} $, $ \ce{O} $, $ \ce{O+} $, $ \ce{N} $, $ \ce{N+} $, $ \ce{e-} $). Under well-established assumptions, the flow in the ICP torch can be considered continuum, partially ionized, and collision-dominated \cite{Magin2004a}. Then, the Navier-Stokes equations are used to express mass, momentum and energy conservation.  The electromagnetic field is modeled with a simplified form of Maxwell’s induction equation, coupled with the momentum and energy equations through Lorentz force and Joule heating effects. As the Reynolds number is typically low ($ Re \sim 100 $), the flow is assumed to be laminar and transition is neglected. The LTE model is adopted, where energy modes are assumed to follow a Maxwell-Boltzmann distribution and equilibrium chemistry occurs. 
Previous work found that the LTE assumption was a reasonable approximation at high pressures within the test chamber \cite{Panesi2007}.
Two sets of simulations were run, i.e.,  including the probe in the domain, and for a free-jet flow, referenced to as CF-ICP-PR and CF-ICP-FJ, respectively.
The simulation requires three main input parameters, specifying the chamber pressure, inlet mass flow rate, and effective electrical power to the coil. The first two values replicate the corresponding experimental measurements, i.e., $\pc$ and $\mdot$. The input electrical power to the simulation assumes a numerical power efficiency, 
$\eta\subsuprm{el}{sim} = \Pelsim/\Pelmeas$, which is typically selected at  $50\%$ \cite{Degrez2001,Viladegut2020}. In the context of this work, we also provide an estimate of this value based on comparison to the OES measurements, as detailed later in the paper.

\begin{figure*}[]
	\centering
	\includegraphics[trim = {0.8cm, 0.0cm, 1.5cm, 0cm},clip,width=.95\textwidth]{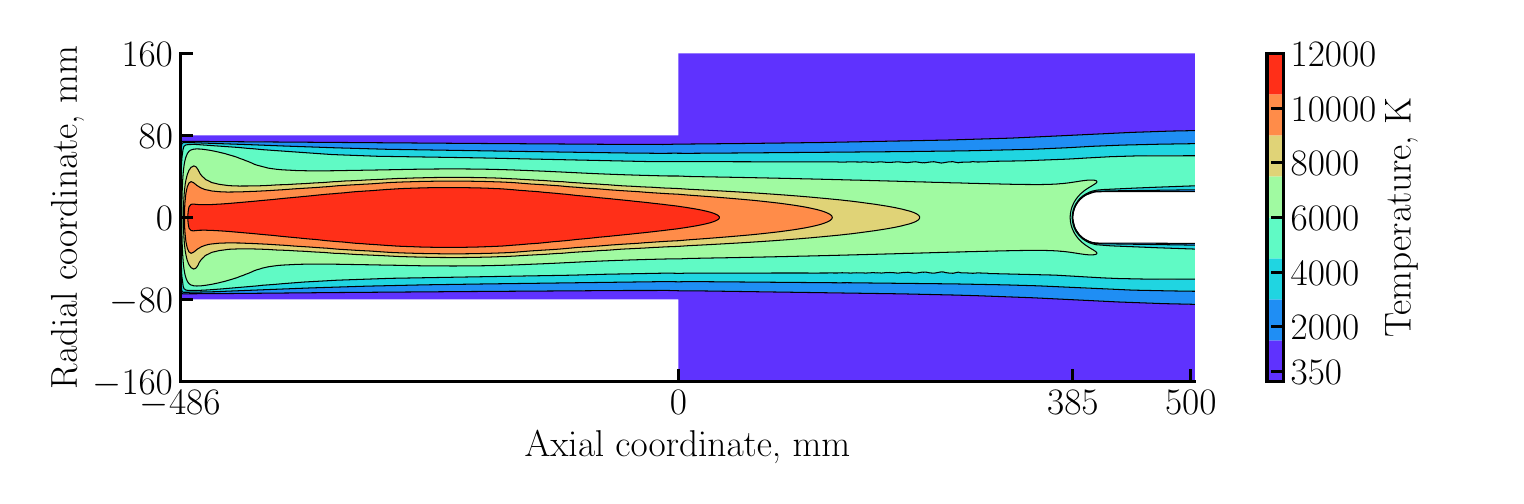}
	\caption[$ \cficppr $ temperature field around an HS50 probe.]{CF-ICP-PR temperature field computed for $ \pc~=~\SI{100}{\mbar} $, $ \mdot = \SI{16}{\gram/\second} $ and $ \Pelsim~=~\SI{100}{\kilo \watt} $ around a HS50 probe at 385~mm from the torch exit.}
	\label{fig:figure2_8}
\end{figure*}

Since equilibrium chemistry is usually not achieved within the BL close to the probe, additional numerical tools, presented shortly after, are needed to refine the computation in this region region and to account for its relevant effect on the wall heat flux. 
In this regard, the BL edge is defined as the position along the stagnation line of the inflection point of the radial velocity gradient, $ \beta~=~\partial v / \partial r $ \cite{Degrez2001}. 
A set of five non-dimensional parameters were defined to characterize the BL edge quantities \cite{Degrez2001}
\begin{equation}
	\begin{cases}
		\Pi_1 = \delta\subrm{e}  / R\\
		\Pi_2 = \beta\subrm{e} \cdot R / u\subrm{t} \\
		\Pi_3 = \parderconst{\beta}{z}{\textrm{e}} \cdot R^2  / u\subrm{t}  \\
		\Pi_4 =  u\subrm{e}  /  u\subrm{t} \\
		\Pi_5 =  u\subrm{e}  / u\subrm{s}, \\
	\end{cases}
	\label{eq:PWT:NDPs}
\end{equation}
where the subscript $ "\textrm{e}" $ indicates the boundary layer edge, $ "\textrm{t}" $ indicates the torch exit point, while $ "\textrm{s}" $ indicates the free-jet point along the axis. 
Fig.~\ref{fig:figure4_3a} shows a schematic of the stagnation line flow impinging onto the probe, highlighting the relevant positions of the measured and computed flow quantities.

\subsection{Boundary layer code}

\begin{figure}[]
	\centering
	\subfigure[]
	{
		\includegraphics[trim = {0cm, 11cm, 19cm, 0cm}, clip, 
		width = 0.55\textwidth]{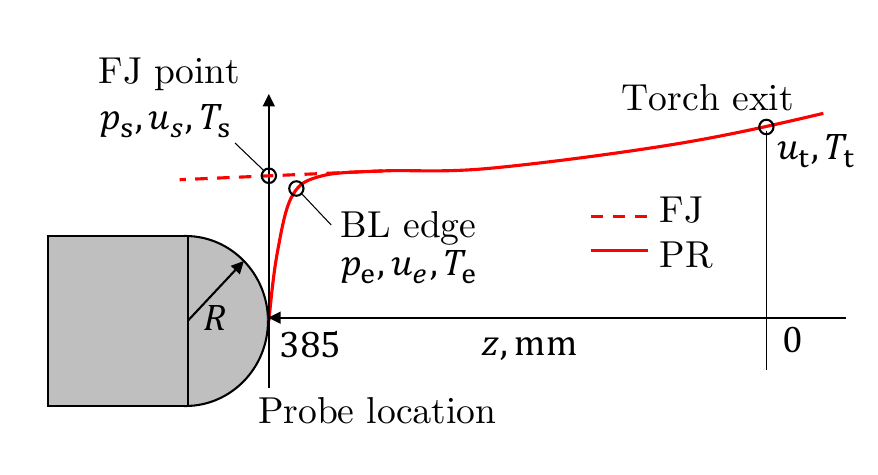}\label{fig:figure4_3a}
	}		
	\subfigure[]
	{	\includegraphics[trim = {0cm, 4cm, 14cm, 0cm}, clip, 
		width = 0.40\textwidth]{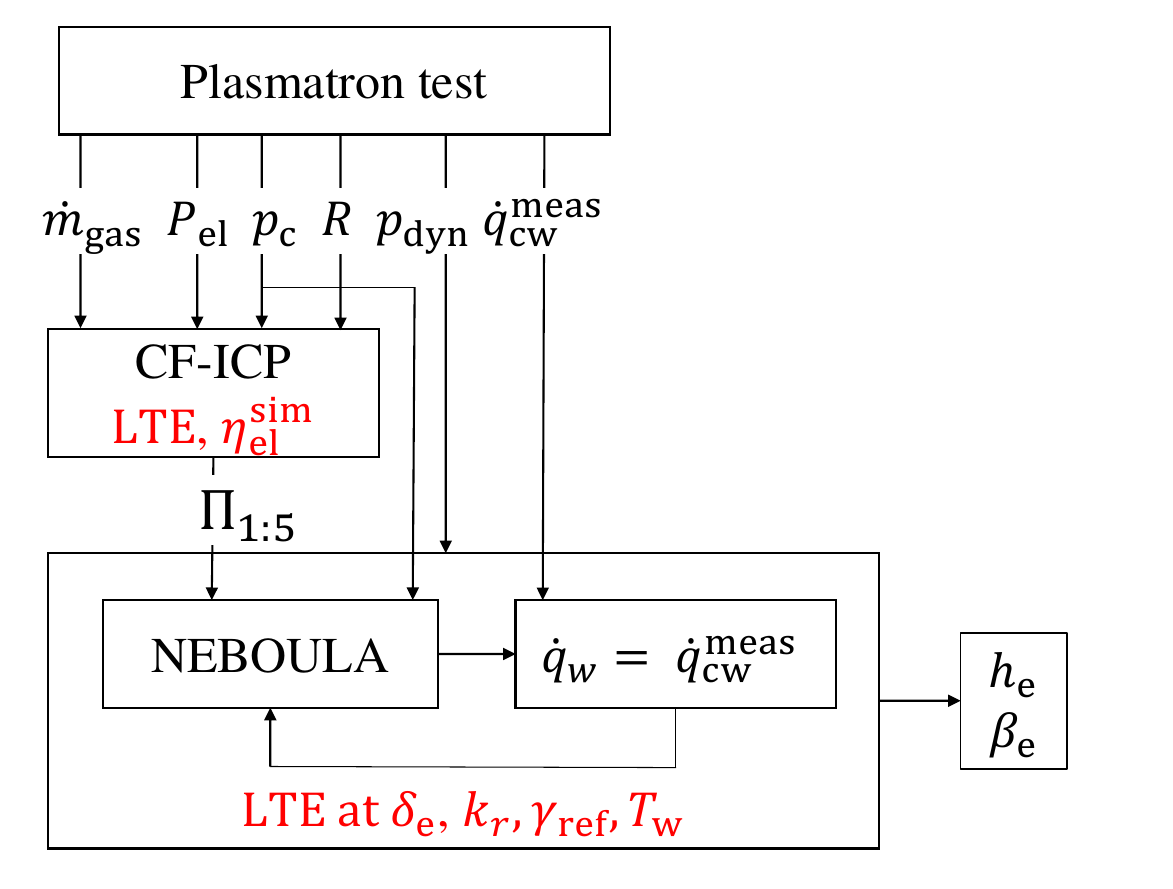}\label{fig:figure2_11}	} 
 	
	\caption[]{(a) Sketch of representative velocity or temperature profiles for flow conditions with the probe (solid line) and free-jet (dashed line), highlighting the main quantities of interest for the following analysis.  (b) Schematic diagram of the inverse procedure for the estimation of the boundary layer edge enthalpy and velocity gradient traditionally adopted for the VKI Plasmatron. The main assumptions are marked in red. }

\end{figure}

The Non-Equilibrium BOUndary LAyer (NEBOULA) code \cite{Barbante2001, Barbante2002, Barbante2006} was employed to compute the Chemical Non-EQuilibrium (CNEQ) BL for the VKI Plasmatron calorimetric probe.
The solver employs an accurate Hermitian-type multipoint finite difference method for computing reacting flows around bodies of revolution. The exact Stefan–Maxwell equations are used to model the diffusion fluxes, and wall catalyticity effects
are accounted for through a set of wall reactions with the associated reaction-rate probabilities.
While low Reynolds numbers ICP flows would require higher order terms in the BL equations, these reduce to the first-order ones around the stagnation region of an axisymmetric body at zero angle of incidence. 
Due to the relatively large thickness of the BL compared to the size of the probe in such conditions, it was suggested to match the BL and external flow fields at the actual position of the BL edge, indicated here as $ \delta\subrm{e} $. 
The external flow is computed with CF-ICP-PR and the BL edge quantities are defined by the non-dimensional parameters in eq.~\ref{eq:PWT:NDPs}. Once the free-jet velocity is estimated from the Pitot measurement, the NDPs are used to evaluate the dimensional quantities and the BL solution is computed. 
The code is employed in an iterative procedure to estimate the BL edge enthalpy and velocity gradient that match the experimental cold-wall heat flux measurement, provided a reference value of the wall catalytic efficiency $\gammaref$, according the schematic depicted in Fig.~\ref{fig:figure2_11}.

\subsection{Stagnation line Navier-Stokes code}

The VKI STAGnation-LINE (STAGLINE) code \cite{Munafo2014}, was used to solve the stagnation line flow upstream of the probe. 
The code implements the conservation form of the stagnation-line governing equations for spherical blunt bodies derived by Klomfass and Muller~\cite{klomfass1997}. The spatial discretization of the equations is based on the cell-centered finite volume method. The numerical convective fluxes can be evaluated by means of a variety
of flux-splitting schemes and second order accuracy in space is achieved by using an upwind reconstruction to obtain the cell interface variables. Both fully implicit and
fully explicit time-integration schemes are implemented
in the code.
The code is coupled to the $\textrm{Mutation}^{++}$ library \cite{Scoggins2020} for the computation of the thermodynamic and transport properties, and chemical source terms within the gas mixture, as well as to solve the gas-surface interaction boundary condition considering a gamma model.
For this particular study, the mesh included 300 cells, with a hyperbolic tangent growth from the wall, where the size of the first cell is kept lower than $ \SI{1}{\um} $. Specific details concerning the input parameters for these simulations will be discussed in detail later in the paper.

\section{Experimental OES results}

OES measurements were primarily taken in compressed atmospheric air (AIRC) at two locations along the jet axis, i.e., at $ 195 $ and $ \SI{385}{\mm} $ from the torch exit. \
Some test conditions were repeated with a synthetic air mixture including 2\% $\ce{H2} $ (AIRH2) to provide $ \ce{H} $ tracing for electron number density measurements. 
Since the LTE $ \Hbeta $ line strength and broadening significantly faint below $ \SI{7000}{\kelvin} $, only powers above $ \SI{210}{\kW} $ were investigated using this mixture.

\begin{figure}
	\centering
	\subfigure[]
	{\includegraphics[trim ={0.5cm, 0.0cm, 3cm, 0cm}, clip,  width=0.99\textwidth]{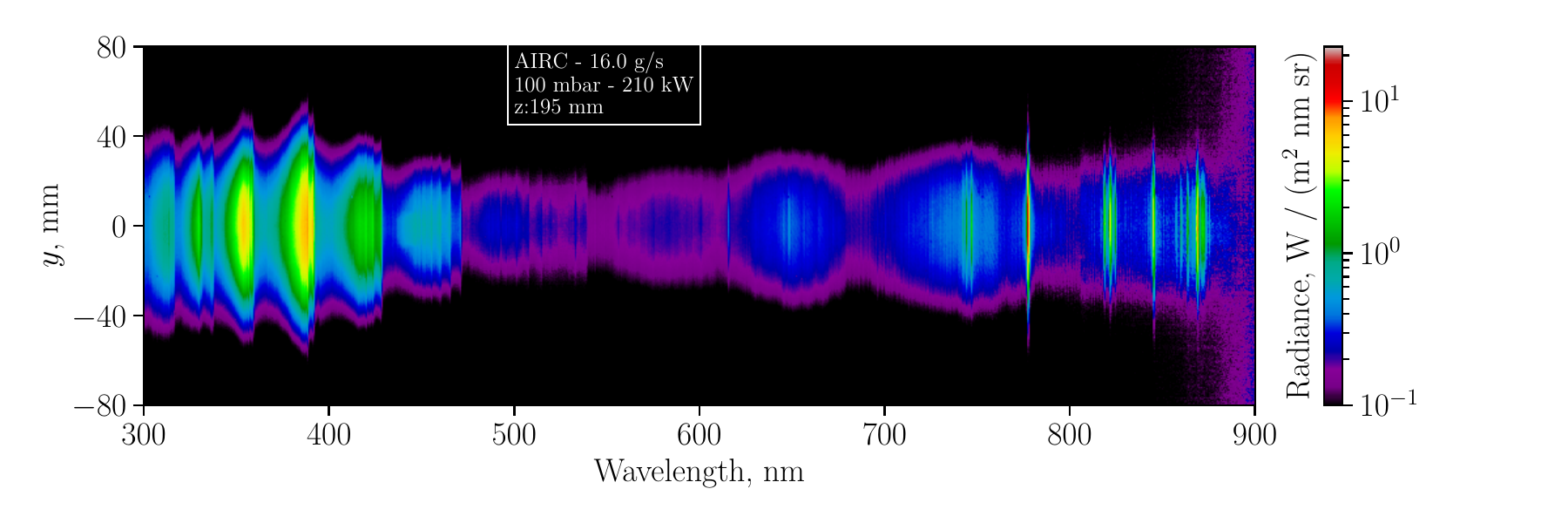}
		\label{fig:figure3_12a}}\\
	\subfigure[]
	{\includegraphics[trim ={0.5cm, 0cm, 3cm, 0cm}, clip,  width=0.99\textwidth]{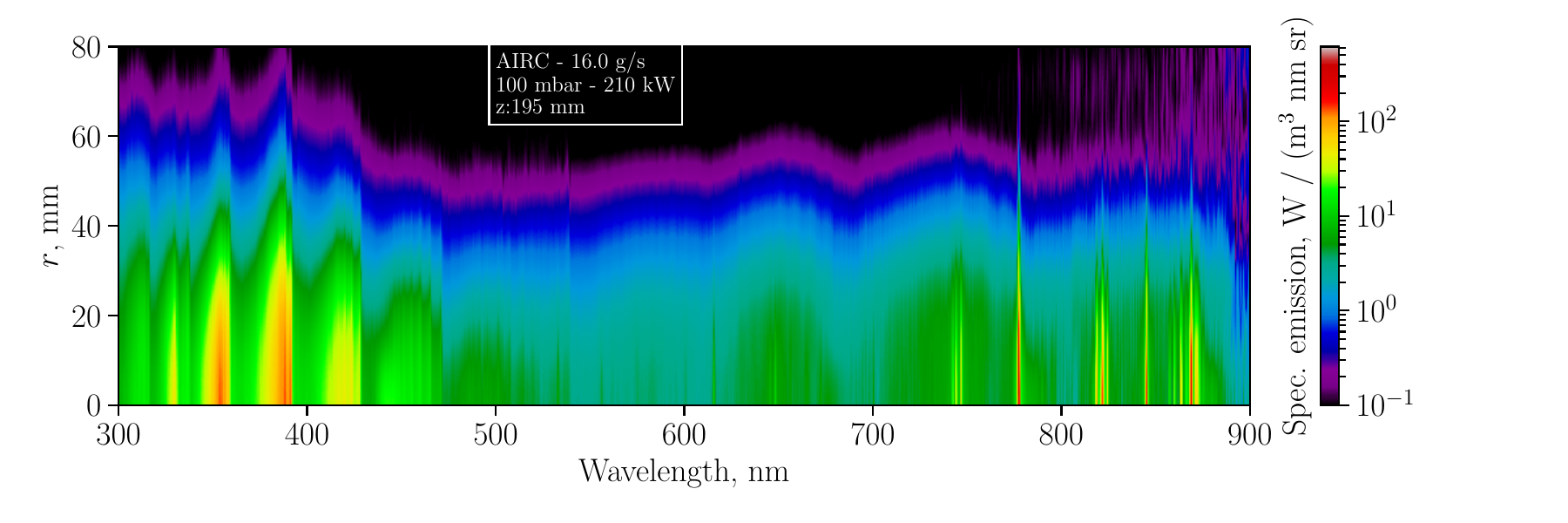}
		\label{fig:figure3_12b}}
	\subfigure[]
	{\includegraphics[trim ={0.cm, 8cm, 0cm, 0cm}, clip,  width=0.99\textwidth]{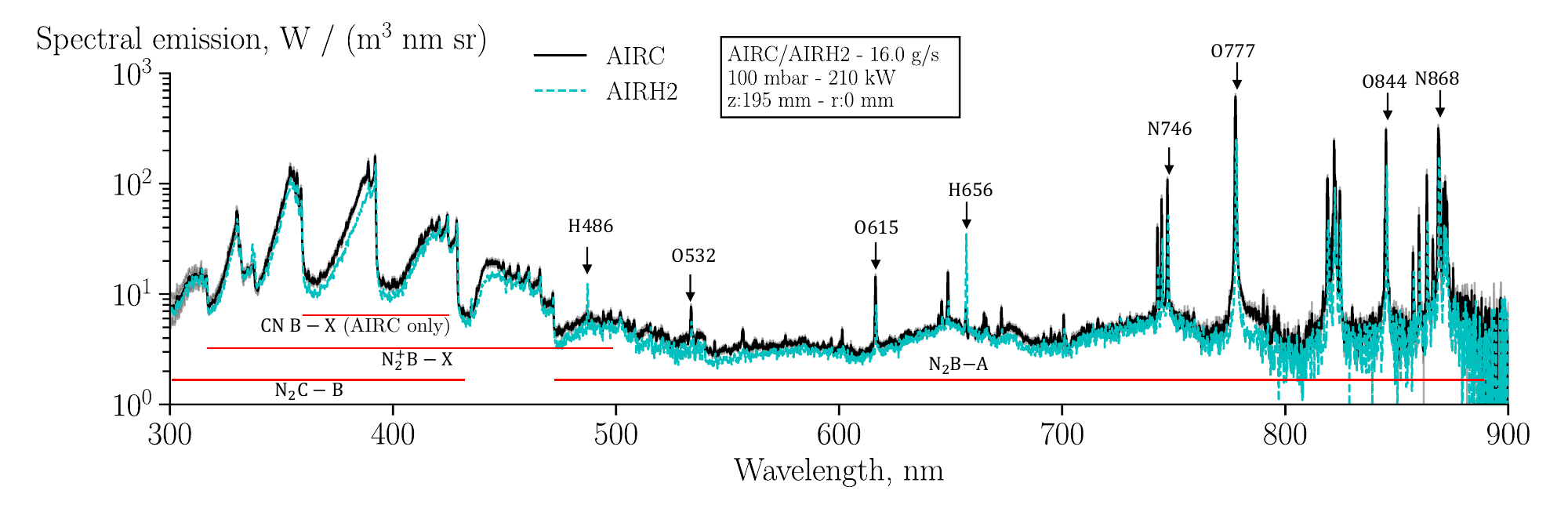}
		\label{fig:figure3_12c}}
	\caption[OES measurement examples.]{Example of measured spatially resolved spectral radiance (a) and rebuilt spectral emission intensity after Abel inversion (b). (c) Comparison between AIRC and AIRH2 measurement at similar conditions, highlighting the main molecular and atomic features (line IDs from table~\ref{tab:APP_OES_atomic_transitions}). }
\end{figure}

Fig.~\ref{fig:figure3_12a} shows an example of calibrated spectral image, measured at $ \pc = \SI{100}{\mbar} $, $ \Pel = \SI{210}{\kW} $ and $ z = \SI{195}{\mm} $ for the AIRC mixture, while Fig.~\ref{fig:figure3_12b} represents the corresponding spectral emission intensity, obtained after Abel inversion. 
Comparison between AIRH2 and AIRC spectra at similar test conditions, reported in Fig.~\ref{fig:figure3_12c}, highlights that the two mixtures provide comparable intensities of the common atomic and molecular features. Differences in the order of 10\% are attributed to the reproducibility of the flow conditions, as well as to the differences in gas composition. Emission of $ \Ntwopfirstneg $, $ \CNvio $ and $ \Ntwosecondpos $  dominate the spectrum below $ \SI{500}{\nm} $. As emission from CN is absent for the AIRH2 mixture, the molecule is likely produced from the $\ce{CO2}$ present in atmospheric air. An ambient $\ce{CO2}$ concentration of 410~ppm was considered to generate the synthetic spectra for the AIRC mixture.   The  $ \Ntwofirstpos $ system is evident above $ \SI{500}{\nm} $, while atomic lines of oxygen and nitrogen populate the spectrum above $ \SI{700}{\nm} $.
The detected signal strongly weakens above $ \SI{880}{\nm} $ mainly due to the loss of efficiency of the spectroscopic system.
The AIRH2 mixture additionally provides significantly stronger intensities of the H486 and H656 multiplets compared to the AIRC mixture, where H is likely to originate from traces of the water vapor in the atmospheric air.

\subsubsection*{Atomic lines}

Using the data listed in table~\ref{tab:APP_OES_atomic_transitions}, we determined the temperature associated to transitions from the electronic excited states. The transition IDs referenced through the following paragraphs are also reported in the table.

\begin{figure}[h!]
	\centering
	\subfigure[]
	{\includegraphics[width = 0.45\textwidth]{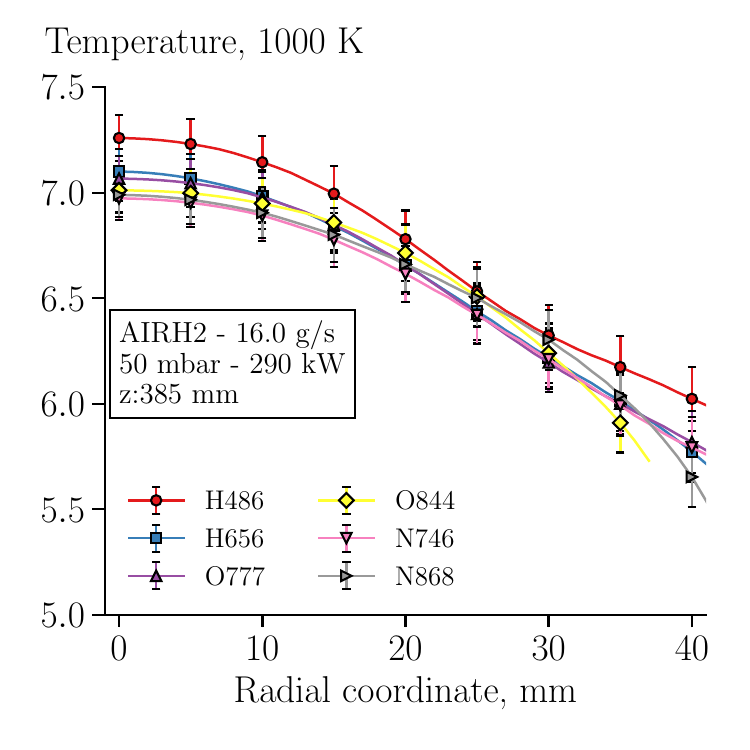}
		\label{fig:figure3_13a}}
	\subfigure[]
	{\includegraphics[width = 0.45\textwidth]{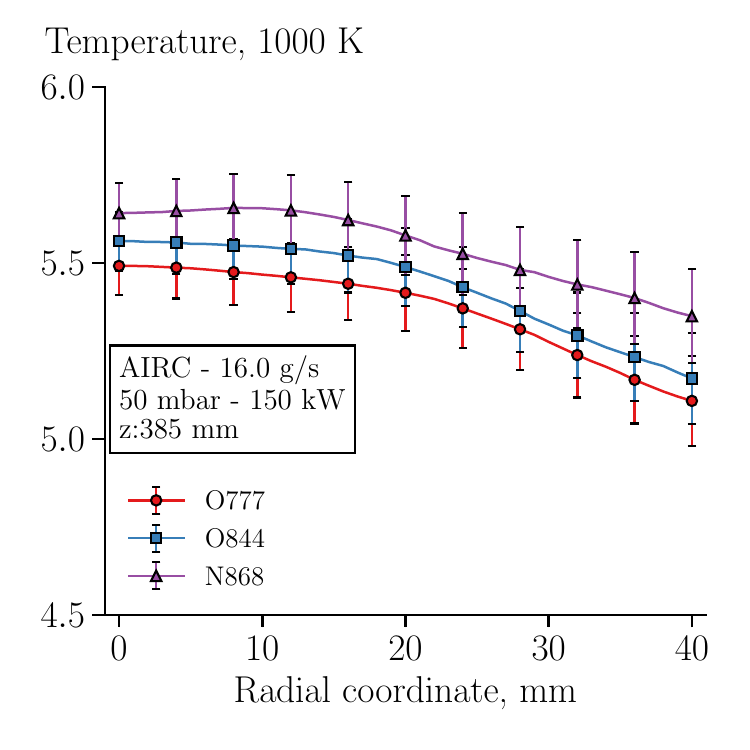}
		\label{fig:figure3_13b}}
	\subfigure[]
	{\includegraphics[width = 0.40\textwidth]{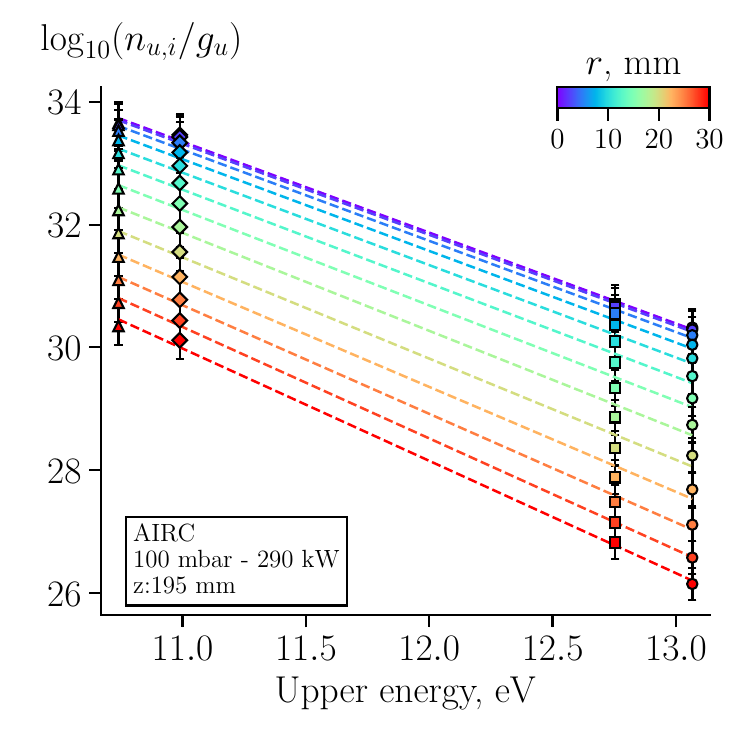}
		\label{fig:figure3_14a}} 
	\subfigure[]
	{\includegraphics[width = 0.40\textwidth]{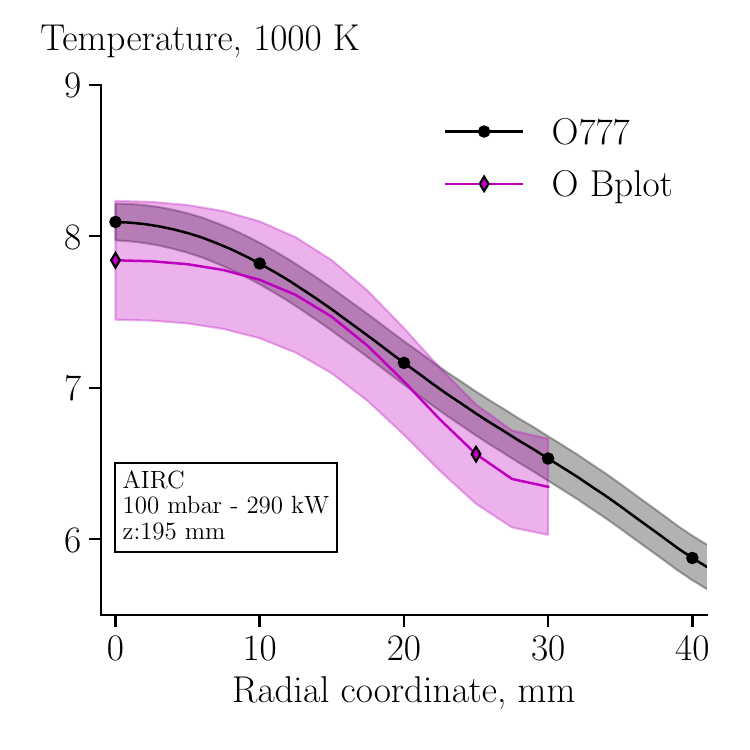}
		\label{fig:figure3_14b}} \\
	\AFmargin{}
	\caption[LTE temperature from Boltzmann plot method.]{(a, b) Example of measured spatial LTE temperature profiles from the atomic lines at different powers for AIRH2 and AIRC mixtures, showing agreement between different species and transitions within uncertainty bounds. (c) Boltzmann plot of the oxygen lines at different radial positions ($ n_{u, i} $ in $ \SI{}{1/\meter \cubed} $), for a selected condition at high power, offering strong detectable lines. (d) Comparison between the temperature measured from the Boltzmann plot method with the one measured from the O777 lines emission. }
\end{figure}

The selected transitions typically offered strong peaks, allowing to resolve the atomic lines above the baseline radiation, mostly due to the $ \Ntwofirstpos $ system, over a large radial extent. 
Fig.~\ref{fig:figure3_13a} and \ref{fig:figure3_13b} show examples of the measured temperature profiles up to $ \SI{40}{\mm} $ from the jet axis at $ z=\SI{385}{\mm} $ and $\pc = \SI{50}{\mbar}$. The former is a high power condition in AIRH2, thus allowing to observe also hydrogen lines, while the latter is a lower power condition in AIRC, limited to oxygen and nitrogen atoms only.
We observe that the temperature profiles from different atomic lines agree within the computed uncertainty bounds. The latter increase from around 1.5\% at the jet centerline to about 2.5\% at $ r =\SI{40}{\mm} $.
The maximum deviation between the nominal values is mostly limited around $ \SI{100}{\kelvin} $ for the strongest lines.
For the AIRH2 case, the H486 line shows a larger  positive deviation, which can be attributed to a significant lower emission intensity that may decrease the detection accuracy.
For the low power condition at $ \SI{150}{\kW} $, instead, only the strongest lines, i.e., O777, O848 and N868, are detectable with sufficient intensity with respect to the background $\ce{N2}$ radiation.
Above $ r = \SI{30}{\mm} $, the signal to baseline ratio is considerably larger for the oxygen lines, mostly due to the lower dissociation temperature of $\ce{O2}$ with respect to $\ce{N2}$, and, hence, higher abundance of the former atoms. 
Moreover, with respect to other O and N lines, the O777 triplet features higher transition probabilities, and better accuracy, thus providing the strongest and most reliable detectable signal across the investigated conditions. As a result, this will be considered as a reference throughout the following analysis for comparison with different techniques.

The electronic temperature was also measured from the Boltzmann plot method of eq.~\ref{eq:boltzmann_plot} applied to the oxygen lines. 
In this case, the O532, O615, O777 and O848 transitions were considered, since the accuracy of this technique increases for larger separation of upper state energy levels. The need for a reliable signal for the O532 and O615 multiplets, however, restricted this application only to the highest power conditions.
Fig.~\ref{fig:figure3_14a} shows the Boltzmann plot of the state densities over the radial extent up to $ r = \SI{30}{\mm} $ from the jet axis for a pressure of $ \SI{100}{\mbar} $, where a linear fit adequately represented their relative distribution within the estimated accuracy.
The intensity of the O532 and O615 lines became significantly weaker after this radial location, preventing sufficient detection accuracy.
Uncertainty is estimated to about ±5\%, due to the large sensitivity to the measured emission intensity, caused by the small energy spacing of the probed excited energy states.
In Fig.~\ref{fig:figure3_14b}  we can observe that the temperature profile obtained from the Boltzmann plot method agree, both in intensity and radial decay, with the LTE values extracted from the O777 line.

Overall, the agreement between temperatures extracted from different lines, as well as the O state densities showing a Boltzmann population, when available, suggest equilibrium of the atomic states and a gas composition close to the one predicted by chemical equilibrium.

\subsubsection*{Spectral fits}
\begin{figure}[h]
	\centering
	\subfigure[]
	{\includegraphics[trim ={0.0cm, 0cm, 0cm, 0cm}, clip,  width=0.99\textwidth]{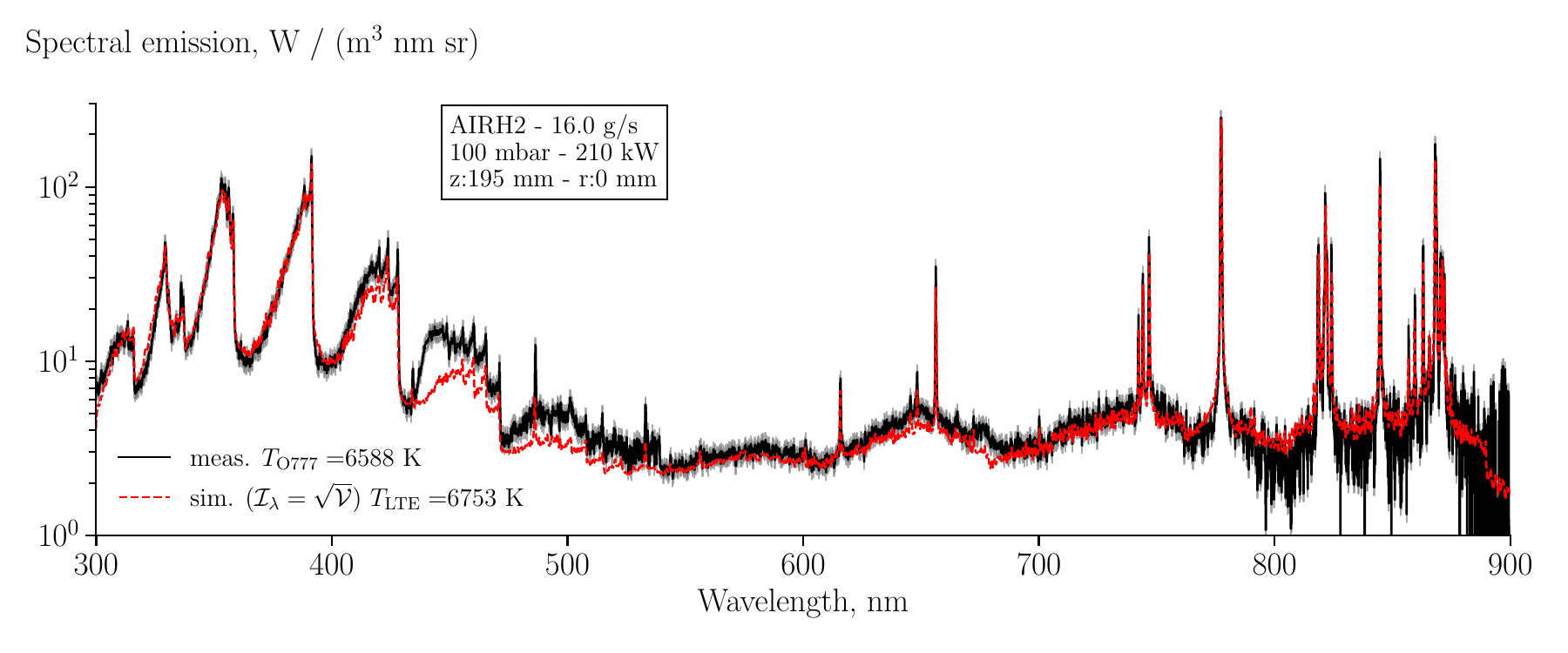} \label{fig:figure3_15a}}

	\subfigure[]
	{\includegraphics[width = 0.99\textwidth]{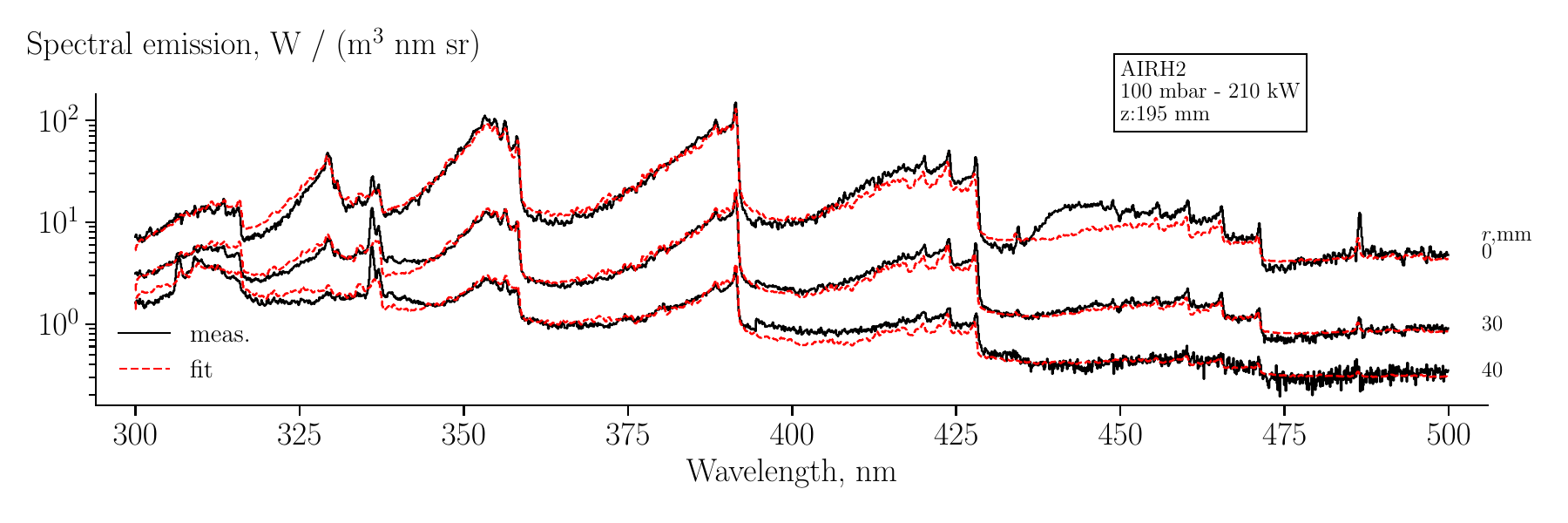}\label{fig:figure3_15b}}

	\caption[Comparison between measured and simulated spectra.]{Comparison between measured and simulated LTE spectral emission intensity for the AIRH2 mixture: (a) UV-NIR range at the jet centerline, (b) UV-visible range at different radial locations.}
\end{figure}

While the previous analysis focused on the integrated line intensities, the measured spectra are examined over a broad range in this section. 
Considering the AIRH2 mixture, for which the mixture composition is accurately known, the LTE synthetic spectra, computed with NEQAIR and convolved with the experimental ILS, are compared to absolute intensity measurements in Fig.~\ref{fig:figure3_15a}.  
The LTE spectrum computed at $ \SI{6753}{\kelvin} $ provides a close agreement over most of the measured wavelength range, resulting in a value about 170~K higher than the one computed from the O777 line at this condition.
While this value exceeds the uncertainty bound estimated on the latter of about $ \pm \SI{100}{\kelvin} $, the residual discrepancy could be due to the presence of background continuum radiation in the experimental data, a phenomenon that was also observed in OES measurements in shock tube facilities \cite{Cruden2009}, or to the accuracy of the experimental ILS.
Agreement is degraded between 430 and $ \SI{530}{\nm} $, where the equilibrium synthetic spectrum, mostly associated to the $ \Ntwopfirstneg $ system, considerably underpredicts the measured spectral emission.
Restricting the observation below 500~nm, Fig.~\ref{fig:figure3_15b} compares experimental measurements and fitted spectra at different radial locations.
Agreement in absolute intensities is confirmed up to $ r = \SI{40}{\mm} $, while the underprediction of the feature between 430 and 530~nm is reduced with increasing values of $ r $.

\begin{figure}[h!]
	\centering
	\subfigure[]
	{\includegraphics[trim ={0.0cm, 0cm, 0cm, 0cm}, clip,  width=0.31\textwidth]{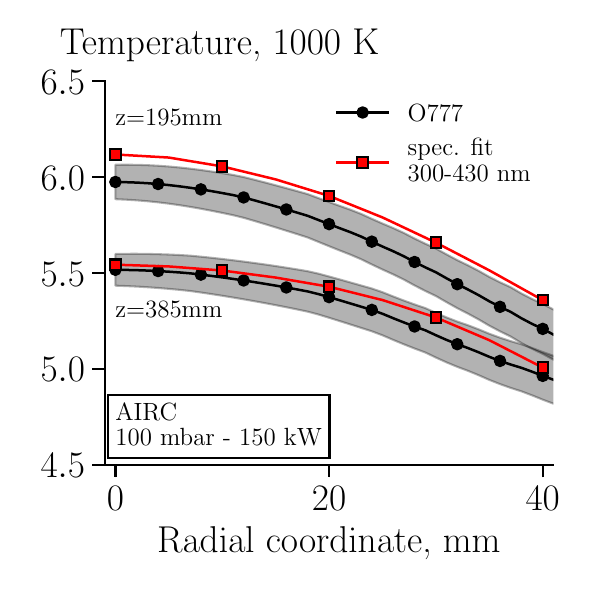}
		\label{fig:figure3_16a}}
	\subfigure[]
	{\includegraphics[trim ={0.0cm, 0cm, 0cm, 0cm}, clip,  width=0.31\textwidth]{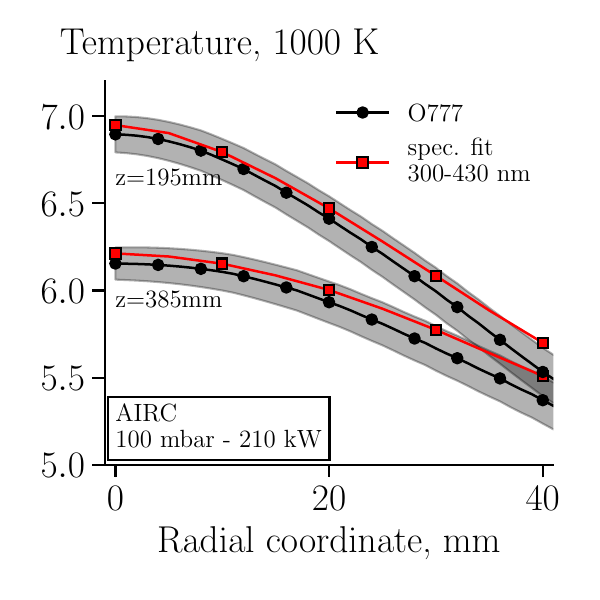}
		\label{fig:figure3_16b}}
	\subfigure[]
	{\includegraphics[trim ={0.0cm, 0cm, 0cm, 0cm}, clip,  width=0.31\textwidth]{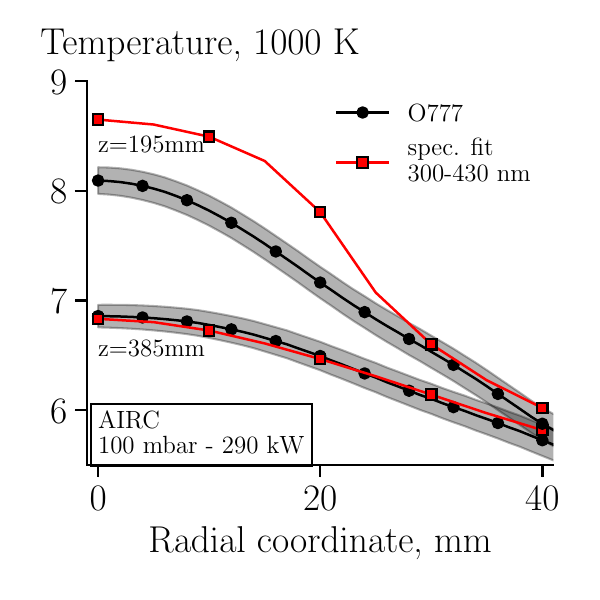}
		\label{fig:figure3_16c}} \\
	\subfigure[]
	{\includegraphics[trim ={0.0cm, 0cm, 0cm, 0cm}, clip,  width=0.31\textwidth]{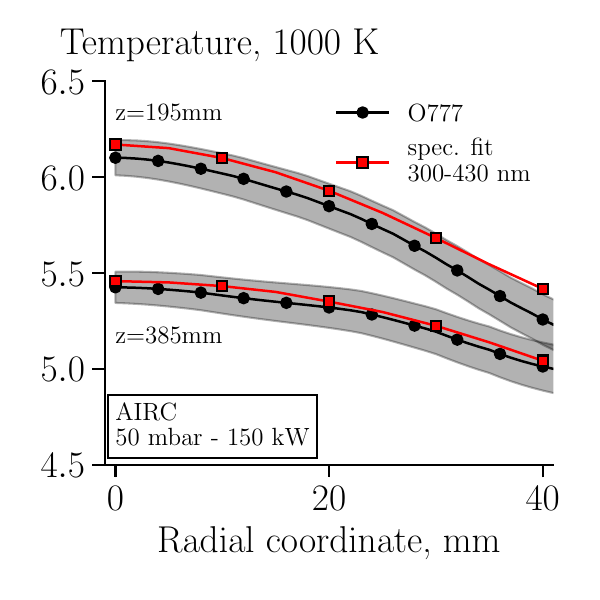}
		\label{fig:figure3_16d}}
	\subfigure[]
	{\includegraphics[trim ={0.0cm, 0cm, 0cm, 0cm}, clip,  width=0.31\textwidth]{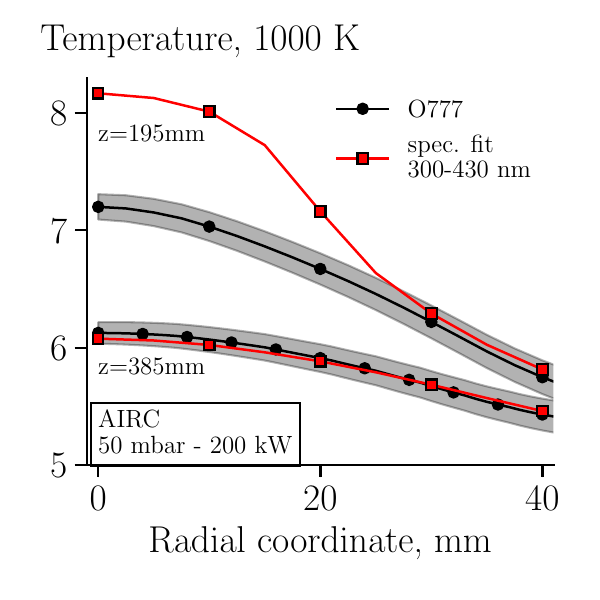}
		\label{fig:figure3_16e}}
	\subfigure[]
	{\includegraphics[trim ={0.0cm, 0cm, 0cm, 0cm}, clip,  width=0.31\textwidth]{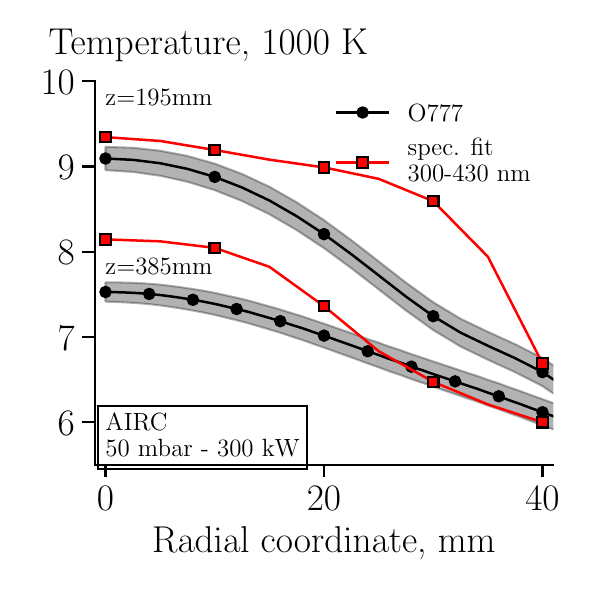}
		\label{fig:figure3_16f}} \\
	
	\caption[Comparison between O777 line temperature and spectral fitting.]{ Comparison between radial temperature profiles obtained from the O777 lines and spectral fits within 300-430~nm, for different chamber pressures and electric powers in AIRC mixture. Agreement is observable at $z=\SI{385}{\milli\meter}$ for temperatures lower than 7000~K.}
	\label{fig:FS2022-AIR-A_O777_T_SF}
\end{figure}

For the purpose of estimating a temperature from the observed molecular features, we restricted the wavelength range between 300-430~nm, in order to avoid a possible bias due to the unrepresented feature.
Fig.~\ref{fig:FS2022-AIR-A_O777_T_SF} compares the temperature profiles estimated from the LTE spectral fits with the results obtained from the reference O777 line across different test conditions for the AIRC mixture.
While the former typically resulted in temperature values about 1 to 2\% larger than the latter, the agreement is confirmed when temperatures do not exceed 7000~K, for both 50 and 100~mbar conditions. 
Above 7000~K, temperatures significantly deviate, and the fit accuracy was largely degraded.
The reason of this mismatch may be related to departures from thermochemical equilibrium that can be expected towards higher input electrical powers, and seem to affect molecular emission in particular.
While this phenomenon is not further investigated in the context of this work, the following analysis will be restricted to a range of temperatures below 7000~K, where the observed agreement supports of the assumption of LTE within the ICP free-jet.

\subsubsection*{Electron number density}

\begin{figure}[]
	\centering
	\subfigure[]
	{\includegraphics[trim ={0.0cm, 0cm, 0cm, 0cm}, clip,  width=0.31\textwidth]{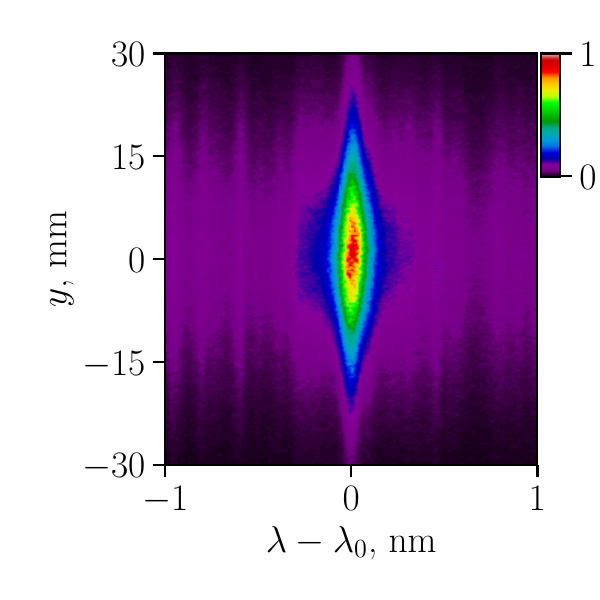}
		\label{fig:figure3_17a}}
	\subfigure[]
	{\includegraphics[trim ={0.0cm, 0cm, 0cm, 0cm}, clip,  width=0.31\textwidth]{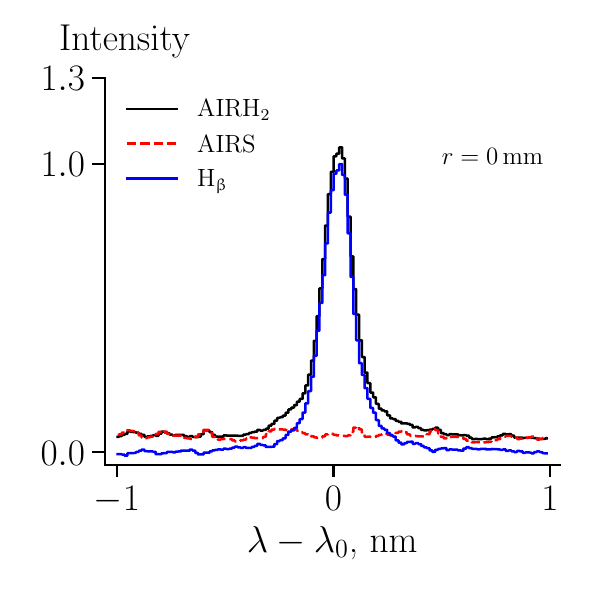}
		\label{fig:figure3_17b}}
	\subfigure[]
	{\includegraphics[trim ={0.0cm, 0cm, 0cm, 0cm}, clip,  width=0.31\textwidth]{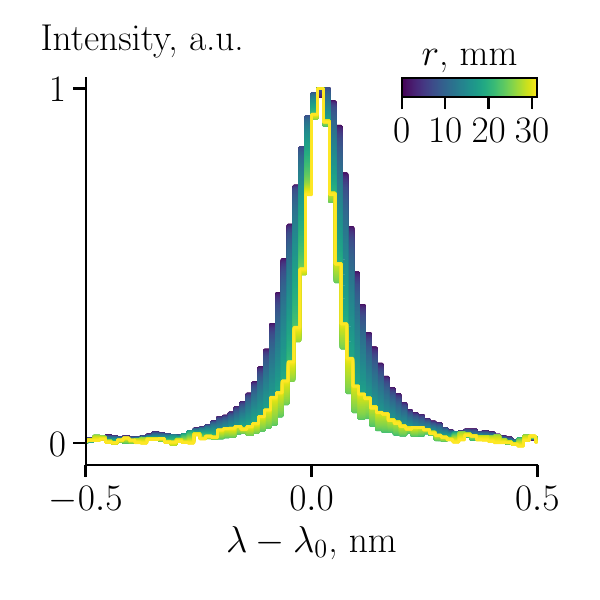}
		\label{fig:figure3_17d}}\\

		\subfigure[]
	{\includegraphics[trim ={0.0cm, 0cm, 0cm, 0cm}, clip,  width=0.35\textwidth]{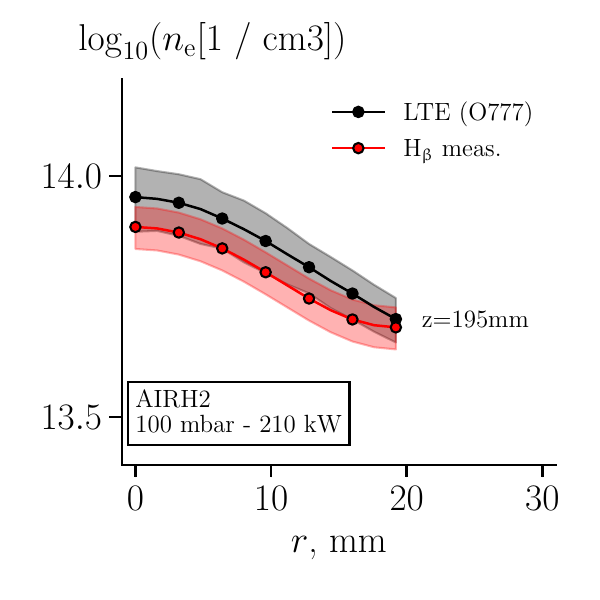}
		\label{fig:figure3_18a}}
	\subfigure[]
	{\includegraphics[trim ={0.0cm, 0cm, 0cm, 0cm}, clip,  width=0.35\textwidth]{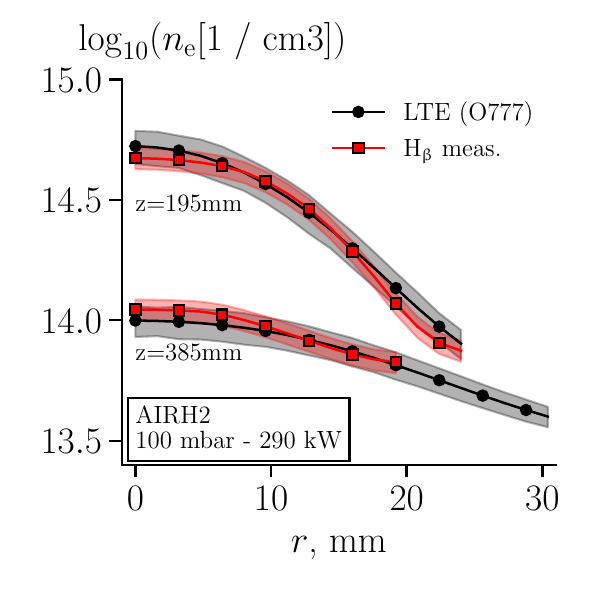}
		\label{fig:figure3_18b}} \\
	
	\subfigure[]
	{\includegraphics[trim ={0.0cm, 0cm, 0cm, 0cm}, clip,  width=0.35\textwidth]{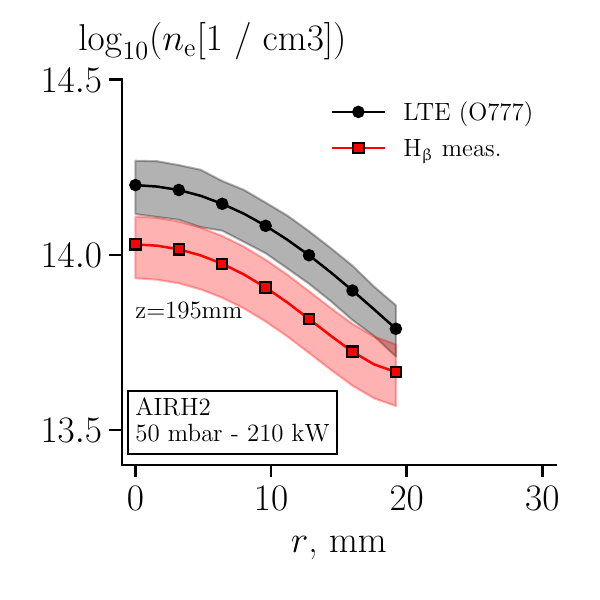}
		\label{fig:figure3_18c}}
	\subfigure[]
	{\includegraphics[trim ={0.0cm, 0cm, 0cm, 0cm}, clip,  width=0.35\textwidth]{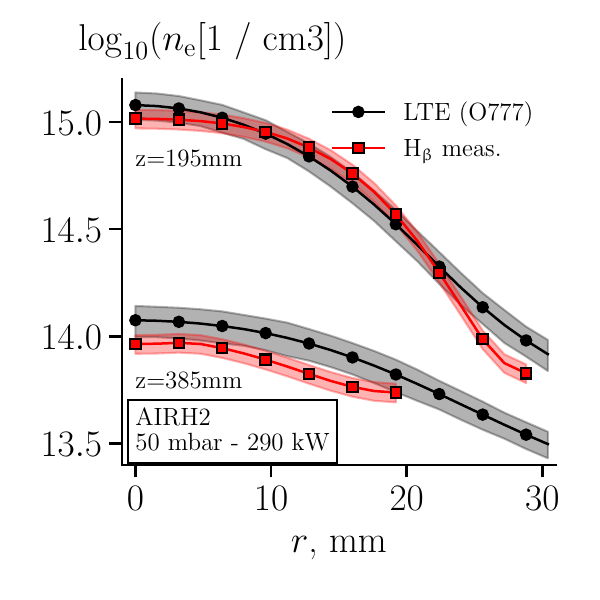}
		\label{fig:figure3_18d}} \\
	
	\caption[]{(a) Normalized spectral image  of the $ \Hbeta $ transition around $ \lambda_0 = \SI{486.1}{\nm} $ for $ \pc = \SI{100}{\mbar} $, $ \Pel = \SI{290}{\kW} $ and $ z = \SI{195}{mm} $. 
		(b) Subtraction of the baseline spectrum measured in synthetic air (AIRS) at $r = \SI{0}{\milli\meter}$.
		(c) $ \Hbeta $ lineshapes as a function of the radial position after Abel inversion and baseline subtraction. (d-g)
		Measured electron number density profiles agree with LTE computations based on the O777 line temperature for the conditions investigated at $ 50 $ and $ \SI{100}{\mbar} $.}
	\label{fig:FS2022-Hbeta-C_Hbeta}		
\end{figure}

The electron number density was measured from the Stark broadening of the $ \Hbeta $ line, using a $ \SI{1200}{grooves/\mm} $ grating to provide sufficient sensitivity to the width of the lineshape with respect to the instrumental broadening.
Fig.~\ref{fig:figure3_17a} shows a sample spectral image of the hydrogen line around $ \SI{486.1}{\nm} $, for a condition at $ \pc = \SI{100}{\mbar} $, $ \Pel = \SI{290}{\kW} $ and $ z = \SI{195}{mm} $, where the intensity is normalized to its maximum value.

To improve the detection accuracy of the lineshape from the baseline radiation due to the $ \Ntwofirstpos $ system, measurements were repeated in similar conditions using a mixture of synthetic air (AIRS) (79\% $ \ce{N2} $, 21\% $ \ce{O2} $), following the procedure described by \citet{Laux2003}.
The baseline intensity was then adjusted in the processing phase to match the measured signal away from the $ \Hbeta $ line center. 
This procedure revealed necessary to compensate for the repeatability of the Plasmatron settings, as well as the slight change in the mixture composition.
Fig.~\ref{fig:figure3_17b} show the measured spectra at $ r = \SI{0}{\milli \meter}$ with and without $ \ce{H} $ tracing, together with the lineshape after baseline subtraction.
Fig.~\ref{fig:figure3_17d} shows the evolution  $ \Hbeta $ lineshape, normalized to its peak intensity, along the radial coordinate, highlighting the decrease in line width with $ r $, associated to the reduced Stark broadening, as free electrons diminish with the decreasing temperature.

Lineshapes are fit with a Voigt profile to increase the accuracy of the measured FWHM to a subpixel level.
For the investigated conditions, Fig.~\ref{fig:FS2022-Hbeta-C_Hbeta}(d-g) compare the measured electron number densities from the $ \Hbeta $ line widths to the LTE computations based on the O777 temperatures. 
Agreement is appreciable in both absolute values and radial decay, additionally supporting that conditions close to equilibrium are likely to be achieved.
A larger discrepancy is found at 50~mbar, 210~kW, for which only the limit uncertainty bounds are overlapping. 
It is relevant to observe that for the high power conditions, i.e., at 290~kW for both 50 and 100~mbar, agreement is observed in atomic line temperatures, electron density measurements, and Boltzmann plot of the oxygen lines, while the LTE spectral fits are deviating.
The reasons for this mismatch are not further investigated within the context of this study, but future work aiming at expanding the OES characterization of the ICP jet towards higher power conditions should consider a more detailed analysis of the molecular bands.

\section{Comparison to CF-ICP and enthalpy rebuilding method}
\label{sec:OES_CFICP_comparison}

We compare the solutions provided by the numerical codes involved in the traditional rebuilding procedure, i.e., CF-ICP and NEBOULA, to the set of experimental data, including OES and intrusive measurements of heat flux and dynamic pressure on hemispherical probes of different size.
The temperature of the free-jet plasma flow at $ z~=~\SI{385}{\mm} $ and $ r~=~\SI{0}{\mm} $, that is, $ T\subrm{s} $, relied on the OES measurement of the oxygen multiplet around $ \SI{777}{\nm} $ (O777), due to the highest signal to noise ratio throughout all the test conditions.

\subsection{Temperature profiles}

We exploited the database of CF-ICP simulations, generated every $ \SI{10}{\kW} $ of input numerical power, for both $ 50 $ and $ \SI{100}{\mbar} $ chamber pressures and $ \SI{16}{\gram / \second} $ mass flow rate, using an eleven-species air mixture. The chamber pressure and inlet mass flow rate to CF-ICP are given by the experimental values. To estimate the effective numerical input power, temperature profiles of free-jet simulations were linearly interpolated between the adjacent values of input power, such that the experimental value at $ z = \SI{385}{\mm} $, $ r = \SI{0}{\mm} $ is matched, following the approach previously presented in Ref.~\cite{Fagnani2020a}.
The numerical power efficiency is computed as $ \eta = P\subsuprm{el}{sim*} / \Pel $, where $ P\subsuprm{el}{sim*}  $ is the linear interpolation between the two selected numerical power inputs that provides the match.

\begin{figure}[h]
	\centering
	\subfigure[]
	{\includegraphics[trim ={0.0cm, 0cm, 0cm, 0cm}, clip,  width=0.31\textwidth]{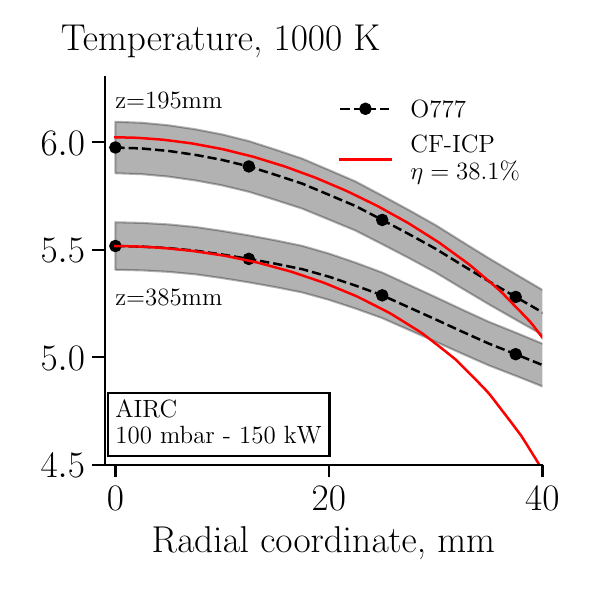}
		\label{fig:figure3_19a}}
	\subfigure[]
	{\includegraphics[trim ={0.0cm, 0cm, 0cm, 0cm}, clip,  width=0.31\textwidth]{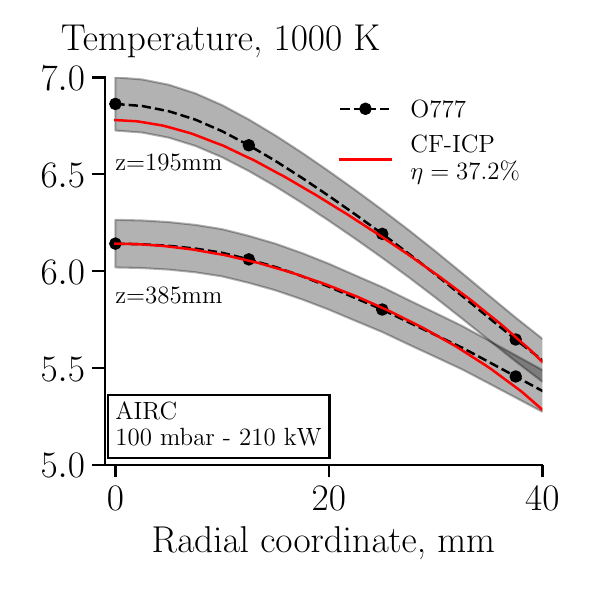}
		\label{fig:figure3_19b}}
	\subfigure[]
	{\includegraphics[trim ={0.0cm, 0cm, 0cm, 0cm}, clip,  width=0.31\textwidth]{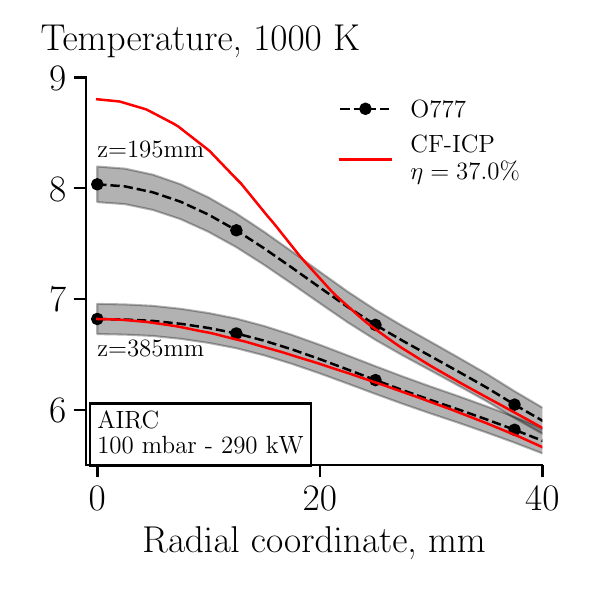}
		\label{fig:figure3_19c}} \\
	
	\subfigure[]
	{\includegraphics[trim ={0.0cm, 0cm, 0cm, 0cm}, clip,  width=0.31\textwidth]{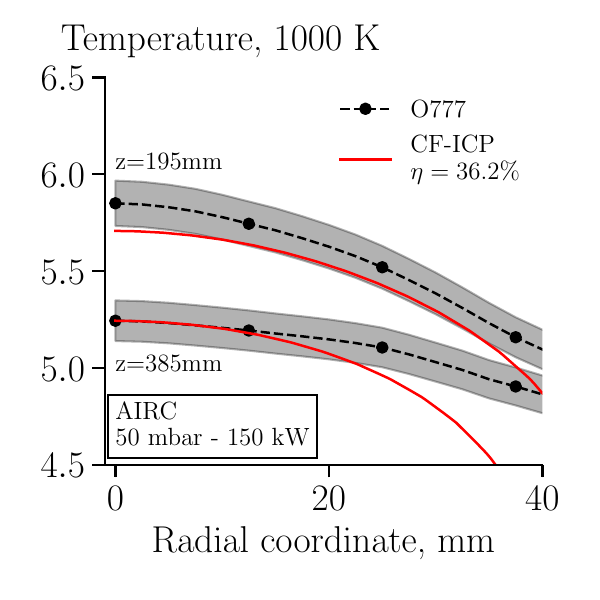}
		\label{fig:figure3_19d}}
	\subfigure[]
	{\includegraphics[trim ={0.0cm, 0cm, 0cm, 0cm}, clip,  width=0.31\textwidth]{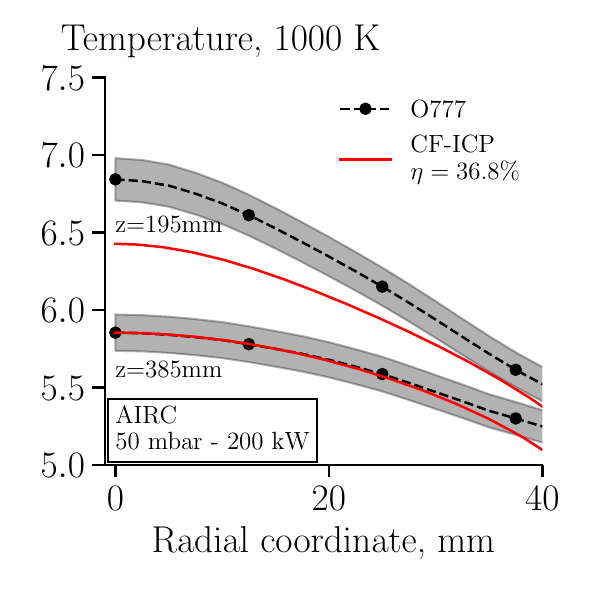}
		\label{fig:figure3_19e}}
	\subfigure[]
	{\includegraphics[trim ={0.0cm, 0cm, 0cm, 0cm}, clip,  width=0.31\textwidth]{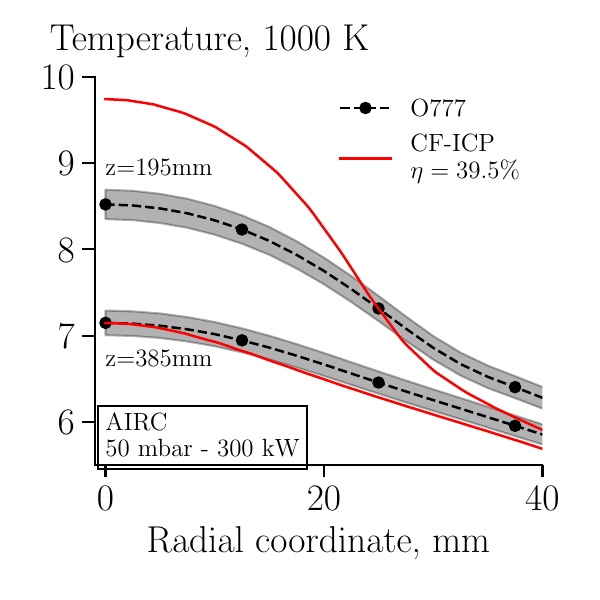}
		\label{fig:figure3_19f}} \\
	
	\caption[Comparison between OES and $ \cficp $ temperature profiles.]{Comparison between the experimental temperature profiles and the ones simulated with CF-ICP-FJ, when the input electric power to the simulation is adjusted to match the measured temperature at $ z = \SI{385}{\mm} $ and $ r = \SI{0}{\mm} $.
	}
	\label{fig:FS2022-AIR-A_O777_ICP}
\end{figure}

Fig.~\ref{fig:figure3_19a}-~\ref{fig:figure3_19f} show the comparison for the different input powers and chamber pressures.
For the $ \SI{150}{\kW} $ and $ \SI{210}{\kW} $ cases at $ \SI{100}{\mbar} $ the agreement is relevant. Both the intensity at the two axial positions and the radial decay in temperature is well represented by the numerical code, although the predicted radial decay is steeper after $ r = \SI{20}{\mm} $.
For $ \SI{50}{\mbar} $ at $ 150 $ and $ \SI{200}{\kW} $, instead, the relative intensity between the two axial locations is under-predicted by the code.

A different situation appears at higher power conditions.
While the radial distribution is well represented at $ z = \SI{385}{\mm} $ at both pressures, the temperature profile at $ z = \SI{195}{\mm} $ is largely over-predicted around the jet axis.
The reasons for this mismatch could be related to modeling assumptions, such as flow steadiness, and physics that the CF-ICP model is not accounting for, e.g., non-equilibrium chemistry and radiation phenomena. 
Experimentally, LTE temperatures from spectral fits were not consistent with the O777 ones above $ \SI{7000}{\kelvin} $, which could indicate departure from equilibrium.

Additionally, the computed numerical power efficiency $ \eta $ is consistent between different experimental conditions, with values between 37 and 38\% at $ \pc =  \SI{100}{\mbar} $ and 36 to 39\% for $ \pc = \SI{50}{\mbar} $.
We should note that these values are lower than $ \eta = 50\% $, traditionally assumed within the VKI enthalpy rebuilding procedure \cite{Degrez2001, Panerai2012a, Viladegut2020}.
While only non-dimensional parameters are extracted from CF-ICP in the context of the latter method, an impact can be associated, e.g., when estimating the BL thickness, thus possibly affecting the heat flux calculation.

We conclude that the CF-ICP can provide a compatible solution with the experimental OES temperature profiles within a range of conditions, provided that the numerical input power is adequately selected.
This would be particularly useful to extend the characterization of the jet away from the stagnation line, thus providing two-dimensional temperature and velocity fields.
However, further comparison with intrusive measurements reveals necessary to corroborate this hypothesis and will be discussed in the next section.

\subsection{Enthalpy - heat flux - dynamic pressure maps}
\label{sec:h_q_p_maps}

Supported by the OES results, we consider the free-jet flow to be close to equilibrium at a distance of 385~mm from the torch exit and for temperatures below 7000~K.
The flow enthalpy is then inferred as $ h\subrm{s} = h\subrm{LTE}(T\subrm{s}, \pc) $, where $\Ts$ is the centerline OES temperature measured from the O777 lines and $ \pc $ is the nominal value of the chamber pressure. An eleven-species air mixture was used, with a composition of $ 79\% ~\ce{N2} $ and $  21\%~\ce{O2} $ by mole. Traces of $ \ce{CO2} $ and $ \ce{Ar} $ negligibly impact the gas enthalpy and were disregarded throughout this analysis.

Uncertainties on the measured $ \Ts $ was estimated about $ \pm 1.5 \% $, and mostly related to the accuracy in the calibration, variability of the measured signal, sensitivity to the Abel inversion and uncertainties in the atomic transition probabilities.
Then, uncertainty on $ \hs $ propagates mainly through the factor $ \partial h/ \partial T $, resulting in maximum values about 6.5\%, while it is negligibly affected  ($ \sim 0.5\% $) by uncertainties on $ \pc $.
Intrusive measurements of cold-wall heat flux were taken on both HS50 and HS30 probe geometries. Accounting also for signal fluctuations, typically observed during measurements, this decreases from about 15\% at $ \SI{0.5}{\mega\watt/\meter\squared} $, to about 7\% above $ \SI{2.5}{\mega\watt/\meter\squared} $. 
The dynamic pressure was measured only on the HS50 geometry, for which we consider uncertainties of $ \pm \SI{5}{\pascal} $.
The Plasmatron electric power $ \Pel $ was simultaneously recorded, for which we assumed a $ \pm \SI{10}{\kW} $ accuracy.
Tables~\ref{tab:APP_REB_exp_data_50mbar}~and~\ref{tab:APP_REB_exp_data_100mbar} in appendix report the numerical values of the experimental data discussed in this section.

\begin{figure}[]
	\centering
	\subfigure[$ \pc = \SI{100}{\mbar} $]
	{\includegraphics[width = 0.43\textwidth]{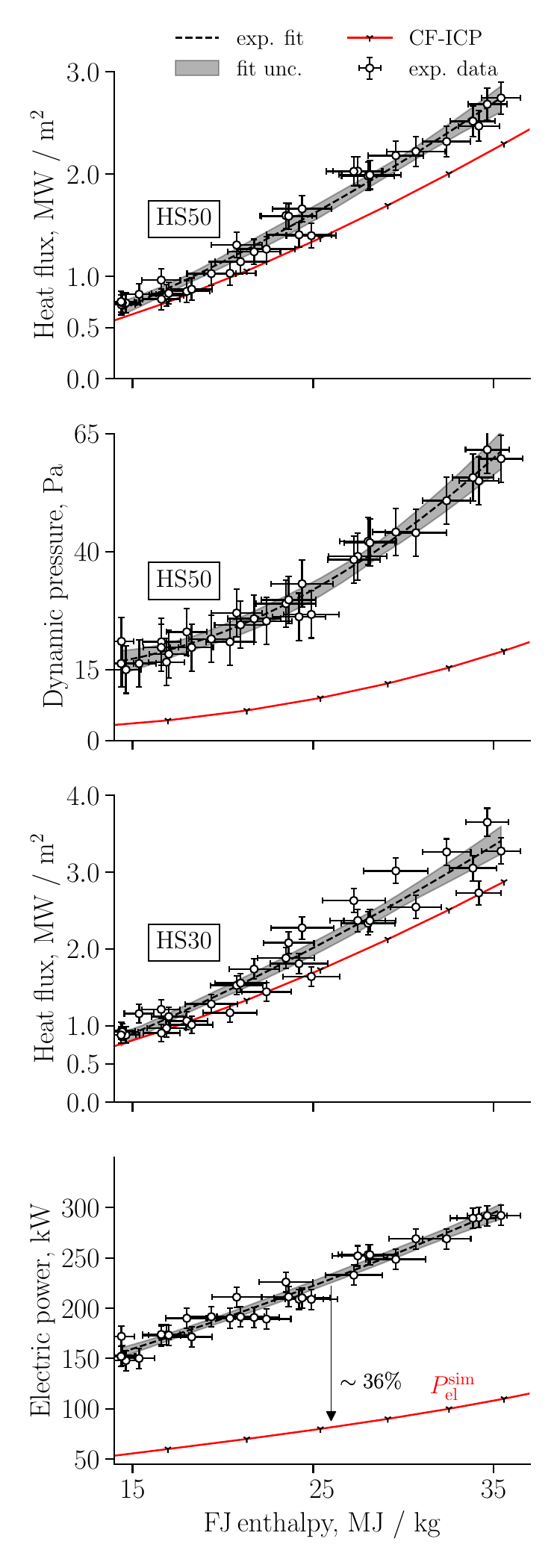}
		\label{fig:figure4_4a}} 
	\subfigure[$ \pc = \SI{50}{\mbar} $]
	{\includegraphics[width = 0.43\textwidth]{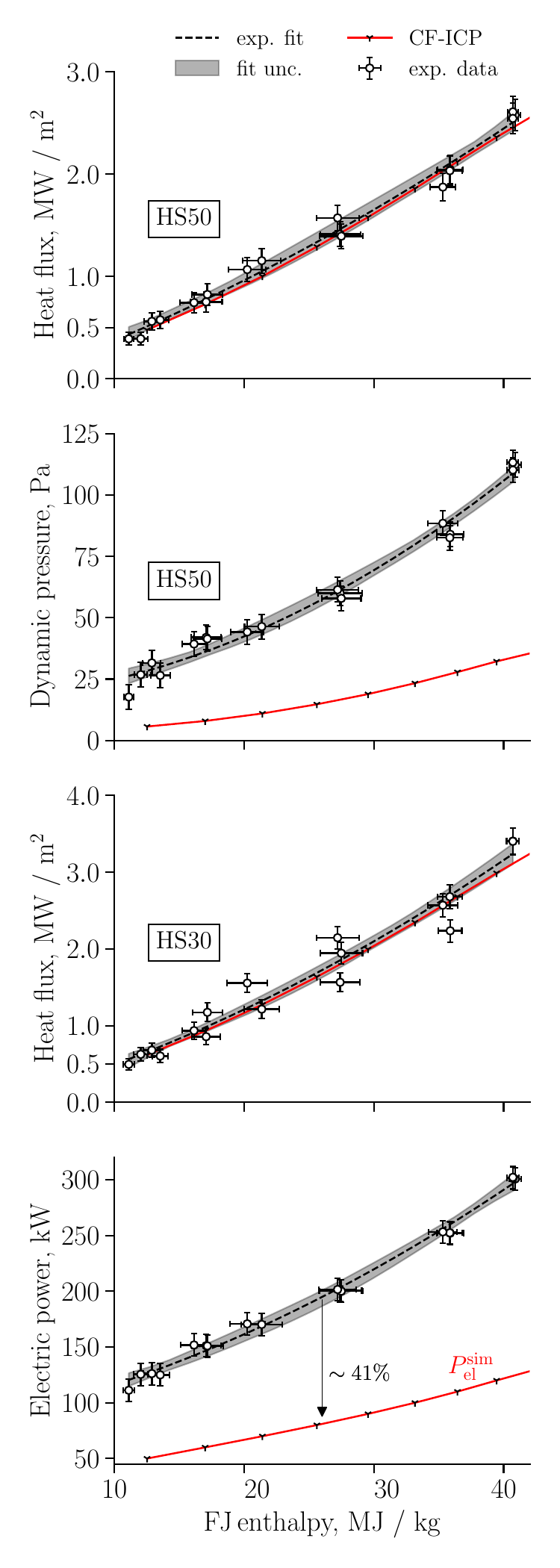}
		\label{fig:figure4_4b}} 
	\caption[Experimental $ \hs\textrm{-}\qcw\textrm{-}\pdyn\textrm{-}\Pel $ maps.]{Experimental cold-wall heat flux (for HS50 and HS30 geometries), dynamic pressure (only for HS50), and electric power, as a function of the measured free-jet enthalpy at (a) $ \SI{100}{\mbar} $ and (b) $ \SI{50}{\mbar} $ chamber pressures. Numerical results from CF-ICP are compared using the free-jet enthalpy as a reference quantity. The ratio of the corresponding CF-ICP numerical power to the measured electric power gives the numerical efficiency indicated close to the arrow in the bottom plots. }
	\label{fig:FS2022-AIR-A_hs_qcw_pdyn_Pel}
\end{figure}

Fig.~\ref{fig:figure4_4a} and \ref{fig:figure4_4b} show the free-jet enthalpy, inferred from OES measurements, versus the measured heat-flux, dynamic pressure and electric power to the facility.
Experimental data are fit with second order polynomials and the shaded gray areas show the uncertainty bounds on these fits. 
Overall, data from different tests provide consistent results and clearly highlight defined trends. 
The scatter around these appears more pronounced for the HS30 geometry, which we relate to the increased sensitivity to the probe position and jet fluctuations, as a result of the smaller probe radius.
Moreover, the heat flux measured on HS30 is consistently larger than the HS50 one for the same value of $ \hs $, which is explained by the smaller radius of the probe, and, hence, larger velocity gradient.

In the same figures, we show the values of heat flux and dynamic pressure computed with CF-ICP-PR, as a function of the corresponding numerical value of the free-jet enthalpy, obtained from a CF-ICP-FJ simulation at the same numerical input power. 
As a result, experimental data and numerical values are compared using the free-jet enthalpy as a common reference quantity. 
We remark that the numerical power efficiency is not considered here.
Rather, as indicated in the bottom plots, this can be estimated from the ratio between the experimental and numerical electric powers at the same $ \hs $. 
Values of $ \sim 36\% $ and $ \sim 41\% $ allow the experimental trend to overlap with the numerical input power for $ \pc = 100 $ and $ \SI{50}{\mbar} $, respectively. 
These values confirm the ones showed previously for some particular conditions in Fig.~\ref{fig:FS2022-AIR-A_O777_ICP}.

While at $ \pc = \SI{100}{\mbar} $ the computed heat flux values similarly underpredict the experimental data both on HS50 and HS30 geometries, at $ \pc = \SI{50}{\mbar} $ they overlap more closely with the experimental trends.
Let us anticipate the following point, which will be demonstrated in \ref{sec:REB_SLFW_validation}: for flow conditions considered here, the heat flux for an equilibrium BL is about the same value of that of a chemical non-equilibrium BL with fully catalytic wall.
As a result, since CF-ICP implements an LTE model, and considering the recombination coefficient of the calorimeter surface likely to be within 0.01 and 0.1, the wall heat flux predicted by  CF-ICP should lie above the experimental data for the same value of $ h\subrm{s} $. 
This discrepancy is attributed to the mismatch in the dynamic pressure, which is significantly under-predicted for both $ 50 $ and $ \SI{100}{\mbar} $ chamber pressures.
This implies that CF-ICP underpredicts the axial flow velocity.
Since the wall heat flux is proportional to the tangential velocity gradient, which, in turn, is proportional to the axial flow velocity, this can explain the underprediction of the former quantity.

In conclusion, while in sec.~\ref{sec:OES_CFICP_comparison} we showed that  CF-ICP could capture approximately well the temperature distribution within the flow, provided that the input power was adequately adjusted to match the free-jet temperature (or, equivalently, the free-jet enthalpy), comparison with intrusive measurements demonstrates incompatible results.
The reasons behind the mismatch in velocity are not further investigated in this work. However, several modeling assumptions within CF-ICP can play a role, including steadiness, laminar flow, equilibrium chemistry, and absence of gas radiation. In this regards, recent advancements in ICP flow simulations, including finite-rate chemistry and unsteady effects \cite{kumar2025}, could lead to a more accurate description of the flow field and provide insights into the observed discrepancies.

\subsection{Comparison with the enthalpy rebuilding procedure}

\begin{figure}[h]
	\centering
	\subfigure[$ \pc = \SI{100}{\mbar} $]
	{\includegraphics[trim = {0cm, 0cm, 11cm, 0cm}, clip, 
		width = 0.48\textwidth]{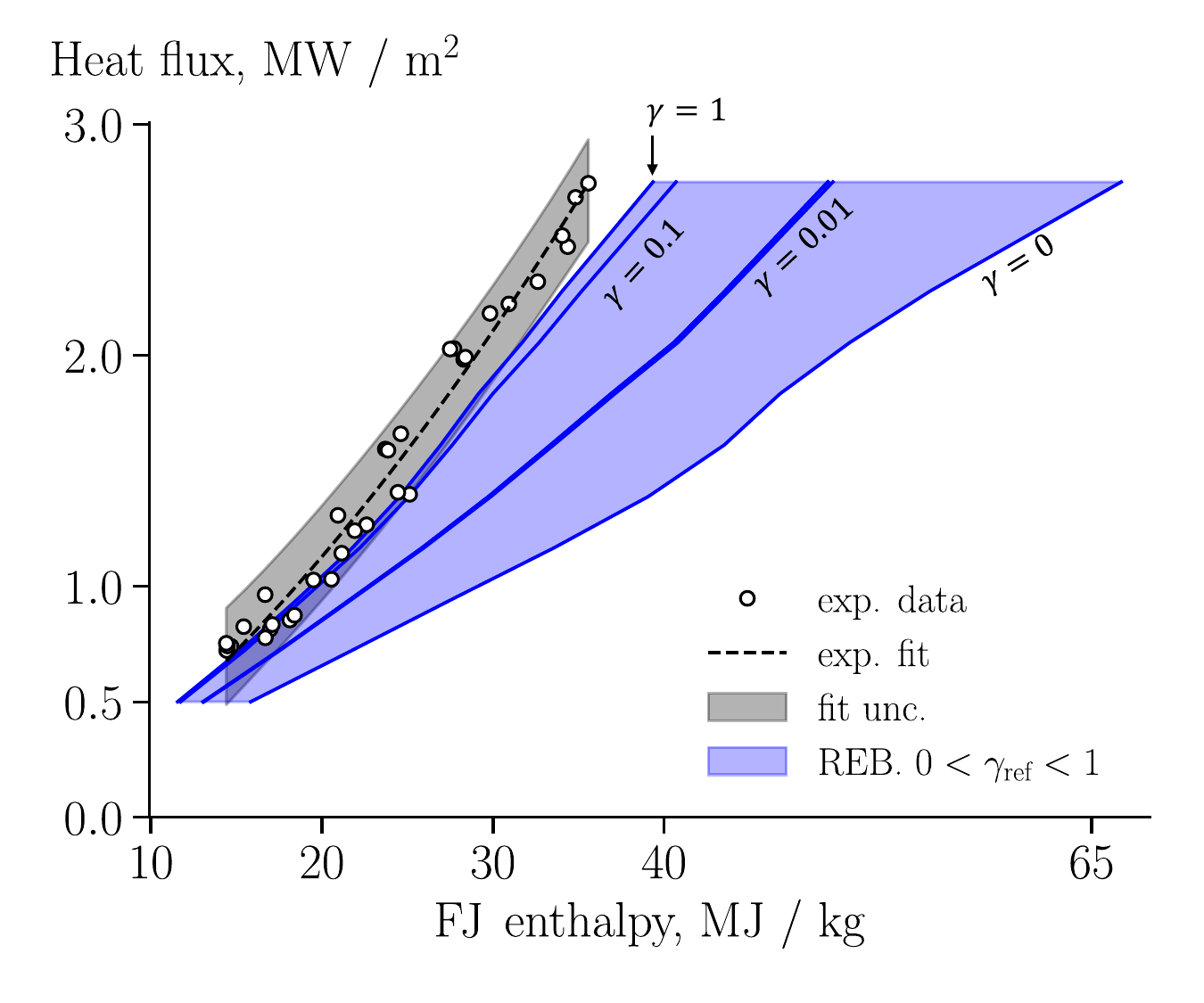}
		\label{fig:figure4_5a}} 
	\subfigure[$ \pc = \SI{50}{\mbar} $]
	{\includegraphics[trim = {0cm, 0cm, 11cm, 0cm}, clip, 
		width = 0.48\textwidth]{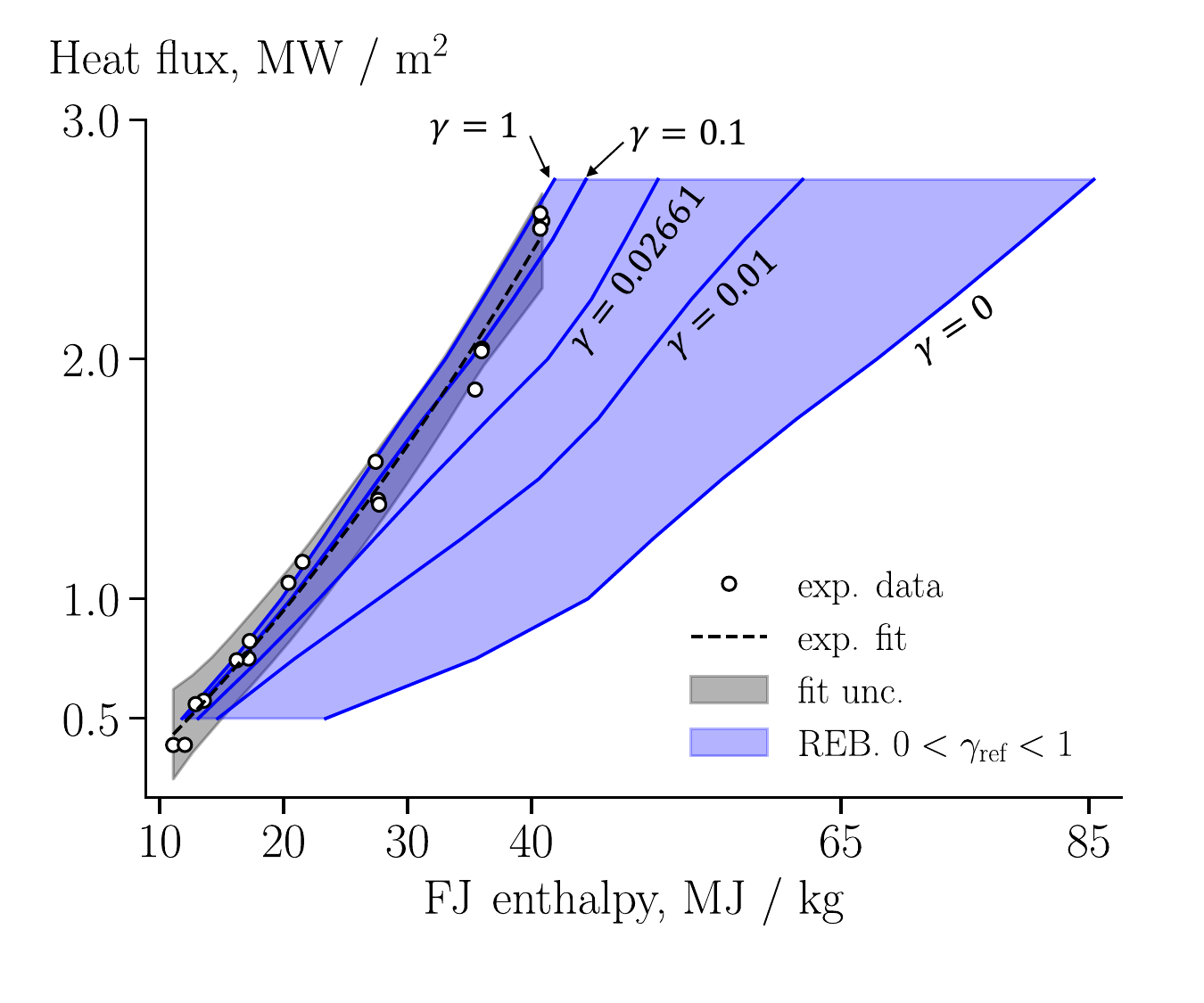}
		\label{fig:figure4_5b}} 
	\caption[Comparison of measured $ \hs\textrm{-}\qcw $ and rebuilt values.]{Comparison between measured free-jet enthalpy vs cold-wall heat flux and rebuilt (REB.) maps: (a) at 100~mbar there is a large discrepancy with the assumption of $ \gammaref= 0.01 $; (b) agreement is found at 50~mbar for $ \gammaref = 0.1 $ \cite{Panerai2012}, while trends diverge for $ \gammaref= 0.02661 $ \cite{Viladegut2020}.}
\end{figure}

The VKI enthalpy rebuilding procedure is compared to the experimental $ h\subrm{s} \textrm{-} \qcw $ maps for the HS50 geometry.
Following the traditional assumptions, a numerical power efficiency of $ \Pelsim / \Pel = \eta\subsuprm{el}{sim} = 50\% $ was used here to compute the NDPs from CF-ICP-PR simulations, selecting the closest available solution in terms of $ \Pelsim $ from the previously generated database, while an eleven-species air mixture with chemical rates from Park \cite{Park2001} was used for the NEBOULA code.
For a consistent comparison to $ h\subrm{s} $, the computed enthalpy at the BL edge, $ \hedge $, was multiplied by the factor $ h\subrm{s} / \hedge $, following the relative trends computed with CF-ICP.
Fig.~\ref{fig:figure4_5a} and \ref{fig:figure4_5b} compare the rebuilt enthalpy envelopes for $ 0~<~\gammaref~<~1 $ with the experimental data. 
Uncertainties on the rebuilt $ \hedge $ are typically around 10\% \cite{Panerai2012, Turchi2017}, without accounting for modeling assumptions.

At $ \pc = \SI{100}{\mbar} $ the reference recombination coefficient traditionally adopted, i.e., $ \gammaref = 0.01 $ \cite{Panerai2012}, provides much higher enthalpies with respect to the ones measured by OES. 
The value suggested by \citet{Viladegut2020} at this pressure, i.e., $ \gammaref = 0.0960 $, produces similar results.
The discrepancy grows from $ \sim 10\% $ at $ \qcw = \SI{0.75}{\mega \watt  / \meter \squared} $ to more than $ 40\% $ at $ \qcw = \SI{2.5}{\mega \watt / \meter \squared} $, thus significantly exceeding the uncertainty bounds and highlighting a diverging trend.
At low heat fluxes the rebuilt envelope shrinks considerably, and the measured trend lies within the computed boundaries. 
At high heat-fluxes, instead, experimental data fall outside the predicted enthalpy-heat flux range, demonstrating incompatibility.

On the other hand, at $ \pc = \SI{50}{\mbar} $, the experimental trend overlaps with the rebuilt enthalpy for $ \gammaref = 0.1 $, which is the value suggested by \citet{Panerai2012} at this pressure.
Considering $ \gammaref = 0.02661 $, as suggested by \citet{Viladegut2020}, instead, departure is noticed above $ \qcw = \SI{1.5}{\mega \watt / \meter \squared} $.
With respect to the $ \SI{100}{\mbar} $ case, experimental data overlap with the $ 0~<~\gamma~<~1 $ envelope within all the measured range, consistently aligning with the higher values of the recombination coefficient.

\section{Consistent numerical simulation of the stagnation line flow}

Motivated by the discrepancies between measured and rebuilt values of the flow enthalpy at $ \SI{100}{\mbar} $, and by the mismatch of CF-ICP in terms of dynamic pressure and equilibrium heat flux limit previously discussed, this section describes an alternative procedure for the numerical computation of the chemical non-equilibrium flow along the stagnation line, seeking compatibility with the entire set of experimental data.
This is desirable for a consistent characterization of the flow, which relies on simulation results to estimate quantities such as the velocity gradient, as well as to understand thermo-chemical phenomena not easily accessible through experiments.

In this case, STAGLINE was used to perform 1D numerical simulations of the flow. The quasi-1D NS equations provide a more accurate model than the BL equations, as they rely on fewer assumptions, and may require fewer input parameters. If the inlet location is selected sufficiently far from the probe to assume that both $ \beta \approxeq 0 $ and $ \beta' \approxeq 0 $ \cite{Turchi2021}, and considering LTE chemical composition at this point, the required boundary conditions are reduced to the free-jet temperature, $ \Ts $, chamber pressure, $ \pc $, and the free-jet velocity, $ \us $.
These quantities can be obtained experimentally, thus avoiding the need of additional numerical parameters.
While $ \Ts $ is inferred by OES, $ \us  $  can be obtained from the measured dynamic pressure. The actual location of the inlet point will be discussed in detail in the next section.
Moreover, STAGLINE allows computing either the subsonic flow in the PWT, and the equivalent hypersonic flight flow-field \cite{Turchi2021}, thus providing a more comprehensive tool overall.

\subsection{Selection of the inlet point location}
\label{sec:delta_star}
As anticipated, the free-jet velocity $ u\subrm{s} $ is obtained from the measured dynamic pressure by means of eq.~\ref{eq:pitot}, accounting for a low-Reynolds number correction through the Homann coefficient, where $ \rhos $ and $ \mus $ are obtained from the experimental $ \Ts $ and $ \pc $ by means of perfect gas and equilibrium relations.
Due to the non-uniform upstream flow, typical of a subsonic jet, the selection of an appropriate inlet location for the 1D simulation is not trivial, and the choice is mainly driven by considerations on the temperature and velocity profiles, which were analyzed with CF-ICP.

\begin{figure*}[h!]
	\centering
	\subfigure[]
	{\includegraphics[trim = {0cm, 0.0cm, 0cm, 0cm},clip,width=.31\textwidth]{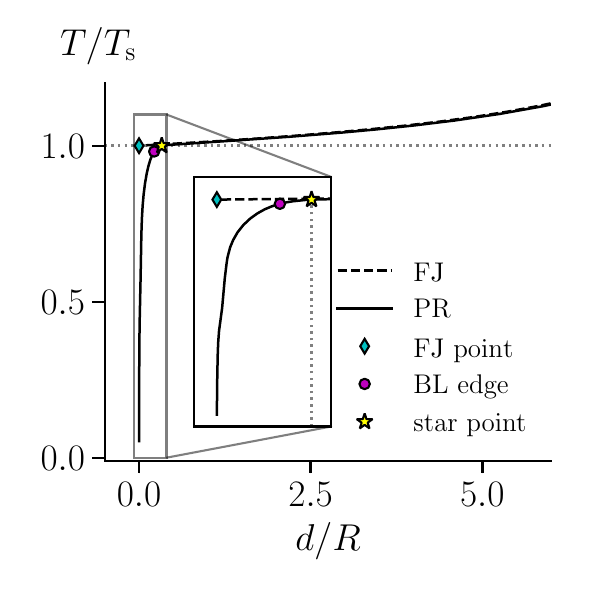}
		\label{fig:figure4_7a}}
	\subfigure[]
	{\includegraphics[trim = {0cm, 0.0cm, 0cm, 0cm},clip,width=.31\textwidth]{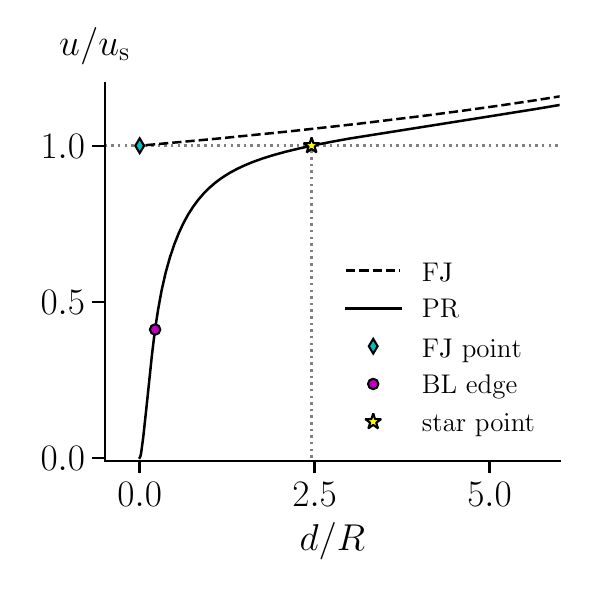}
		\label{fig:figure4_7c}}
	\subfigure[]
	{\includegraphics[trim = {0cm, 0.0cm, 0cm, 0cm},clip,width=.31\textwidth]{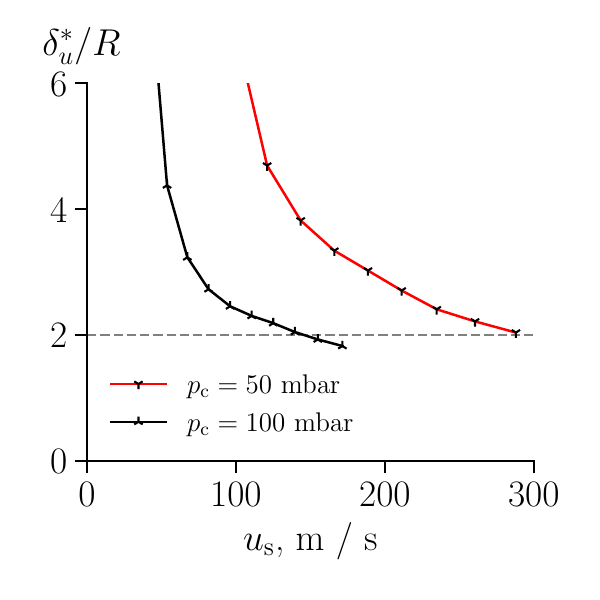}
		\label{fig:figure4_7d}} \\
	\caption[Upstram location of the free-jet quantities.]{Example of temperature (a) and velocity (b) profiles as a function of the distance from the probe along the stagnation line from CF-ICP. (c) Location of the $ \deltastar_{u} $ point, as a function of the corresponding free-jet velocity $\us$, as predicted by CF-ICP.}
\end{figure*}

On the one hand, the influence of the probe on the temperature field of the jet is confined within the thermal boundary layer, typically in the order of $ 5 $ to $ \SI{10}{\mm} $ for the HS50 geometry,  as shown in Fig.~\ref{fig:figure4_7a}.  
We indicate with $ \deltastar_{T} $ the distance from the stagnation point where the axial temperature profile upstream of the probe equals the free-jet point value, $ \Ts $. 
$ \deltastar_{T} / R $ was found to decrease from 0.8 at $ \Ts = \SI{5000}{\kelvin} $ to approach 0.3 for $ \Ts > \SI{7000}{\kelvin} $, almost independently of $ \pc $.
On the other hand, the probe affects the velocity field much further upstream due to pressure perturbations, as it is clearly shown in Fig.~\ref{fig:figure4_7c}. 
We indicate with $ \deltastar_u $ the distance from the stagnation point where the axial velocity profile upstream of the probe equals the free-jet point value, $ \us $.
We found that $ \deltastar_u $ is several times the probe radius, decreasing with $ \us $, and changing appreciably with $ \pc $, as depicted in Fig.~\ref{fig:figure4_7d}.

Then, considering that $ \deltastar_T < \deltastar_u $, and that the location of the inlet point should be unique, this will be defined at $ \deltastar_u $. Moreover, as $\deltastar_{u} $ is typically larger than $2R$, the tangential velocity gradient and its derivative also approach zero at this location, thus providing the convenient approximation of $\beta \approx 0$ and $\beta' \approx 0$. As a result of the quasi-1D formulation, the temperature imposed at $\deltastar_u$ will be nearly uniform downstream, up to $\deltastar_T$.
In practice, once the free-jet velocity is obtained from the measured dynamic pressure, the correlation $ \deltastar_u(\us) $  in Fig.~\ref{fig:figure4_7d} can be used to obtain the corresponding inlet point location. A numerical verification of the procedure against CF-ICP is discussed in \ref{sec:REB_SLFW_validation}.

This procedure inevitably approximates the real flow field for two main reasons.
First, in Sec.~\ref{sec:h_q_p_maps} we concluded that CF-ICP underpredicts the axial velocity for a certain value of the free-jet enthalpy and, hence, $\deltastar_{u}$ can similarly be mispredicted using such correlation. However, this is considered here to be the best approximation with the available computational tools, allowing to capture the inviscid deceleration upstream of the probe, and to avoid the jet decay, which is not modeled in the quasi-1D formulation.
Second, we expect the quasi-1D results to be more accurate for high velocities, as the smaller $ \deltastar_u $ would reduce the influence of the jet decay effects downstream of this location.

\subsection{Application to the experimental conditions}

The procedure is applied to the experimental data reported previously in Fig.~\ref{fig:FS2022-AIR-A_hs_qcw_pdyn_Pel}, for both $ \pc = 100 $ and $ \SI{50}{\mbar} $, and for both HS50 and HS30 probes.
For a value of $ \Ts $ and $\pdyn$, the free-jet velocity is computed according to eq.~\ref{eq:pitot}. 
Then, the position of $ \deltastar_{u} = \deltastar_{u}(\us) $ is obtained from the correlation provided by CF-ICP in Fig.~\ref{fig:figure4_7d}.
Boundary conditions, in terms of $ \Ts $ and $ \us $, are imposed at $ \deltastar_{u} $ and the flow is computed for $ 0~<~\gammaref~<1 $ on the surface, where equal recombination probabilities are assumed for $ \ce{O} $ and $ \ce{N} $. 

\begin{figure}[]
	\centering
	\subfigure[]
	{\includegraphics[trim = {0cm, 0cm, 12cm, 0cm}, clip, 
		width = 0.40\textwidth]{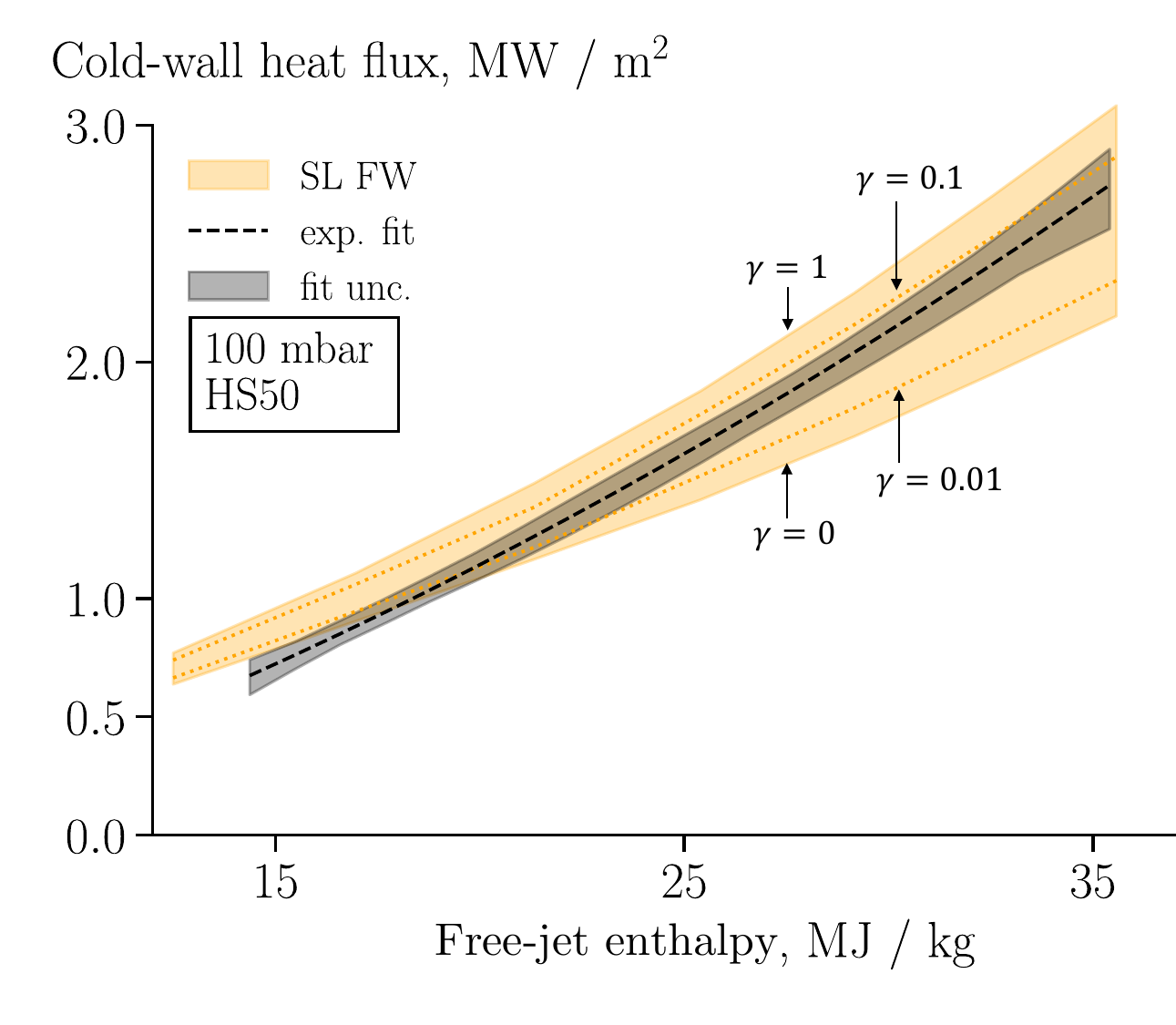}
		\label{fig:figure4_9a}} 
	\subfigure[]
	{\includegraphics[trim = {0cm, 0cm, 12cm, 0cm}, clip, 
		width = 0.40\textwidth]{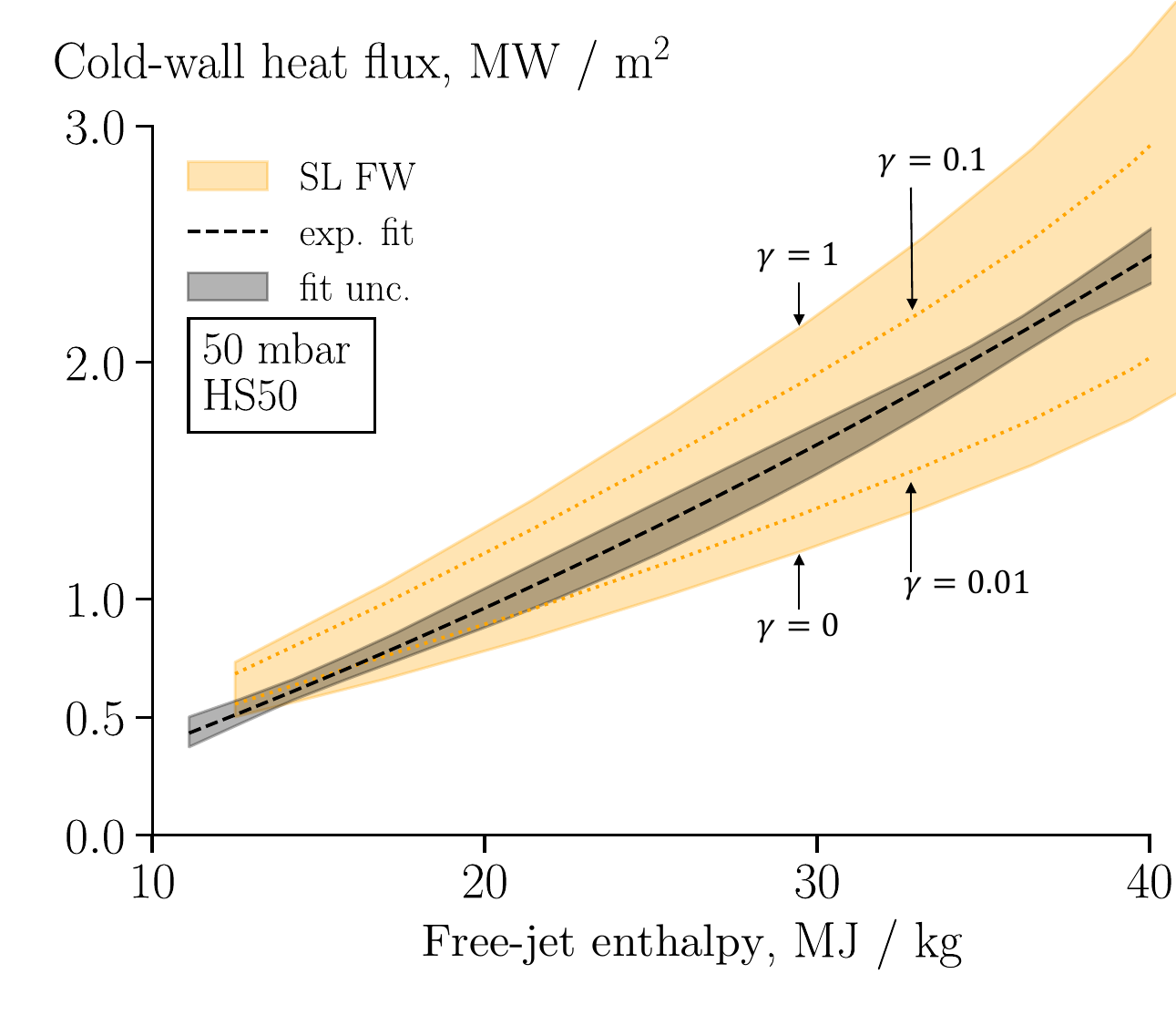}
		\label{fig:figure4_9b}} 
	\subfigure[]
	{\includegraphics[trim = {0cm, 0cm, 12cm, 0cm}, clip, 
		width = 0.40\textwidth]{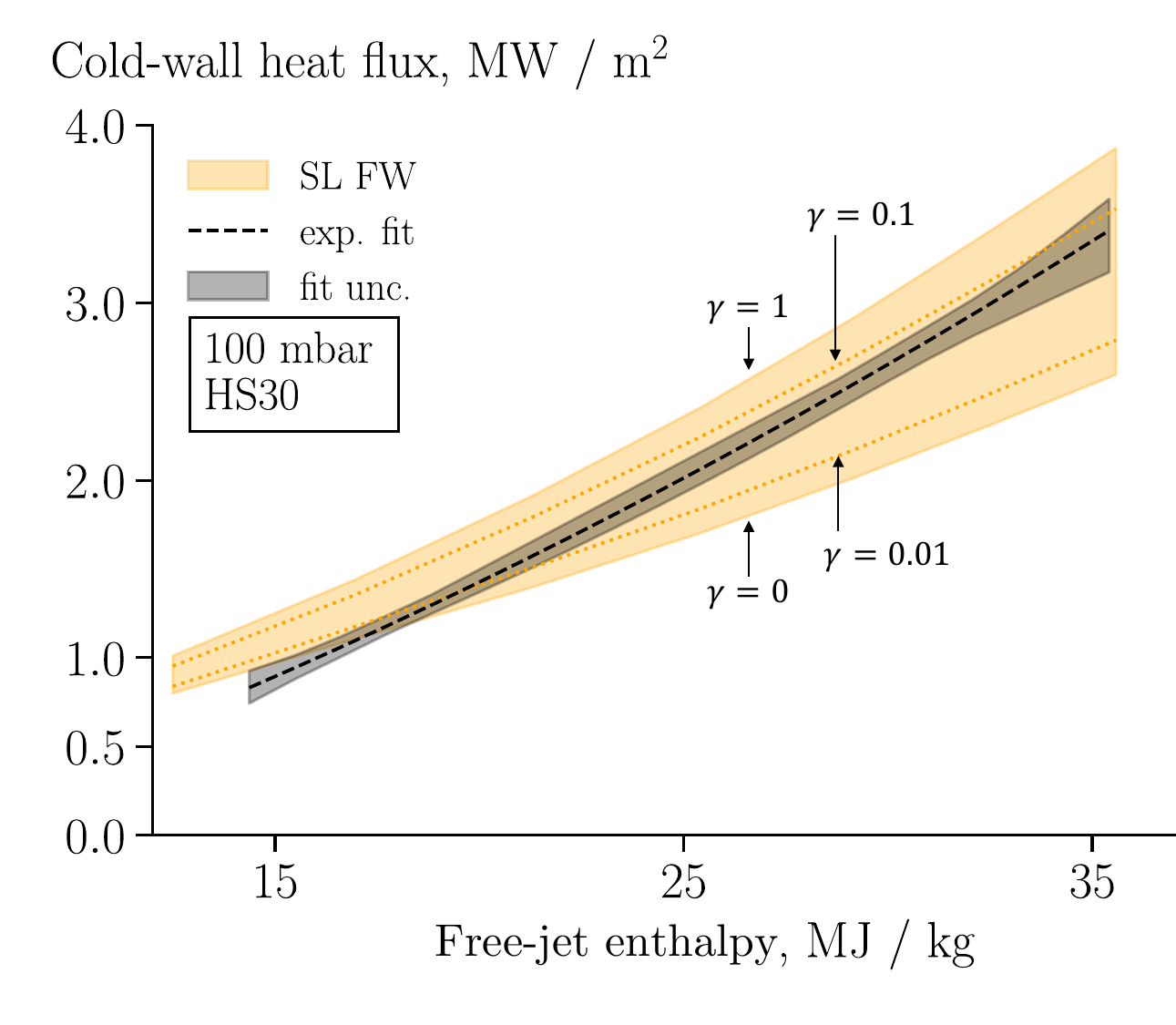}
		\label{fig:figure4_9c}} 
	\subfigure[]
	{\includegraphics[trim = {0cm, 0cm, 12cm, 0cm}, clip, 
		width = 0.40\textwidth]{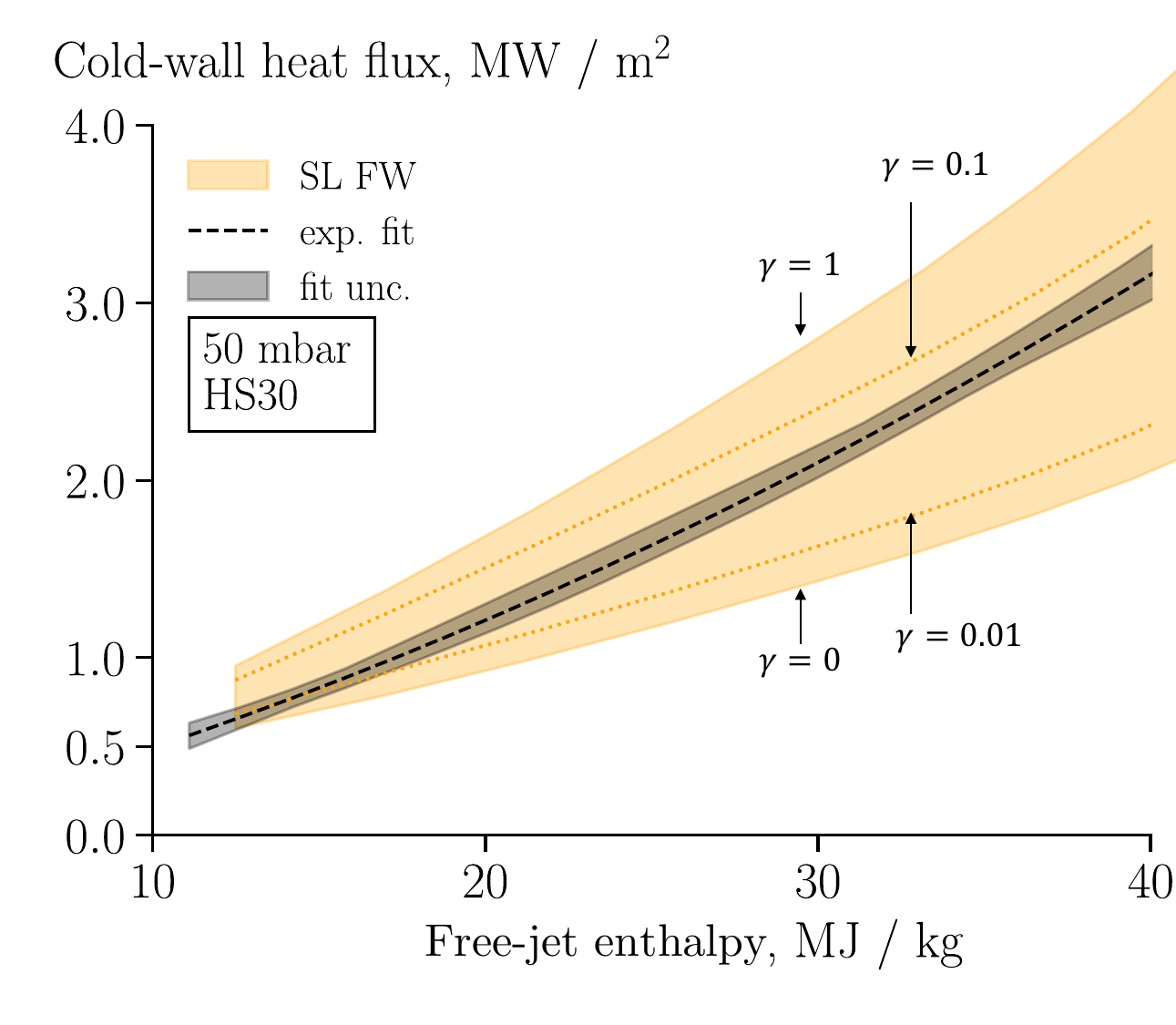}
		\label{fig:figure4_9d}} 
	\caption[Compatibility of new free-jet procedure with measured $ \qcw $.]{Measured $ \hs-\qcw $ trends lie within the $ 0<\gammaref<1 $ envelope predicted with the new procedure (SL FW) starting from the measured $ \Ts $ and $ \us $ at $ \deltastar_{u} $, except below $ \SI{20}{\mega\joule/\kilogram} $ at 100~mbar. The relevance of jet decay effects is believed to degrade the accuracy at low free-jet enthalpies.  }
	\label{fig:FS2022-AIR-A_hs_qcw_exp_SL_FW_EXP}
\end{figure}

Fig.~\ref{fig:FS2022-AIR-A_hs_qcw_exp_SL_FW_EXP} compares the numerical heat flux envelopes, computed with the proposed procedure (SL FW), to the experimental trends, both at 100 and 50~mbar and for the HS50 and HS30 geometries, respectively.
At $ \pc~=~\SI{100}{\mbar} $, simulation results are compatible with experimental data above $ \hs > \SI{20}{\mega\joule / \kilogram} $. At $ \pc~=~\SI{50}{\mbar} $, instead, the compatibility is verified throughout the whole range.
Moreover, the experimental data mostly lie within the predicted $ 0.01~<~\gammaref~<~0.1 $ range, suggesting further compatibility with literature values for the catalytic efficiency of $\ce{CuO}$, previously discussed in Sec.~\ref{sec:h_methods}.  
Additionally, a similar trend is observed for both HS50 and HS30 geometries, providing a solid validation to the proposed procedure.

Concerning the degraded compatibility below $ \SI{20}{\mega\joule / \kilogram} $ at $ \pc = \SI{100}{\mbar} $, the reason can be attributed to the quasi 1D-flow approximation within the procedure. This was anticipated to be poorer for low values of free-jet velocity, as those encountered for low $ \hs $, since the inlet point moves further upstream of the probe and jet decay effects become more relevant. 
Moreover, the predicted heat flux range considerably shrinks for higher chamber pressures and low enthalpies, due to the preponderance of the chemistry within the gas phase, contributing to a limited overlap range.

\section{Conclusions}

This paper presented a detailed analysis of the experimental-numerical methodology employed for the characterization of the subsonic ICP jet in the VKI Plasmatron facility, as well as a new procedure to seek consistency between the set of experimental measurements, including optical emission spectroscopy, and intrusive measurements of heat flux and dynamic pressure, with the numerical simulation results employed to model the flow field.  

Spatially-resolved absolute OES measurements of the free-jet air plasma flow provided radial temperature and electron density profiles under relevant test conditions for PWT material response studies.
Measurements were found compatible with thermo-chemical equilibrium for pressures of $ 50 $ and $ \SI{100}{\mbar} $ and temperatures below 7000~K, at an axial distance of $ \SI{385}{\mm} $ from the torch exit, i.e., the position where the probes were located in the jet within this work, 
while LTE spectral fits departed from the measured emission above $ \SI{7000}{\kelvin} $ and towards $ \SI{195}{\mm} $ from the torch exit.

The selected equilibrium conditions allowed us to infer the gas flow enthalpy from the measured temperature, providing experimental maps of free-jet enthalpy, cold-wall heat flux, and dynamic pressure.
Comparison to CF-ICP showed agreement with the measured radial temperature profiles, provided that the numerical input power was adequately adjusted, while the jet dynamic pressure, and, as a consequence, both the axial flow velocity and stagnation point heat flux, were underpredicted.
The enthalpy rebuilding procedure, based on the inverse solution of the BL equations, and considering typical reference values for the catalytic efficiency assumed for the copper probe, overpredicted the gas enthalpy at $ \pc = \SI{100}{\mbar} $, while results were more consistent with experimental data at $ \pc = \SI{50}{\mbar} $.

These discrepancies were mostly overcome through a different simulation strategy, which solved the quasi-1D Navier-Stokes equations along the stagnation line, using the experimentally measured values of free-jet temperature and velocity imposed as inlet conditions.
In this case, the predicted heat flux was compatible with the experimental data for free-jet enthalpies between 20 and $\SI{40}{\mega\joule/\kilogram}$, and for both 50 and 30~mm diameter hemispherical probe geometries, mostly lying within $ 0.01 < \gammaref < 0.1 $, in agreement with literature data on the catalytic efficiency of $ \ce{CuO} $.
The radial velocity gradient at the BL edge showed a small sensitivity to the wall catalytic efficiency, thus providing a solid framework for the LHTS.
However, agreement was degraded for enthalpies lower than $ \SI{20}{\mega \joule / \kilogram} $ at $ \pc = \SI{100}{\mbar} $, and mainly attributed to the quasi one-dimensional approximation employed.

This work provides a step towards a more consistent framework for the characterization of the ICP flow in PWTs, allowing improved definition of the test parameters, as well as input conditions for material response simulations.
While preserving the LTE assumption, in this case supported by OES diagnostics, the proposed methodology provides significant advantages, being independent of uncertainties in the choice of the catalytic efficiency of the reference probe, as well as of the reaction rates and numerical parameters required by a traditional inverse heat transfer approach.

Further work is advised to improve the free-jet flow velocity measurements, as well as to investigate the departure from the predicted equilibrium emission at high-power conditions.
Extension to both lower pressures and higher powers will be required to fully characterize achievable test envelope.
Improved modeling efforts within the CF-ICP framework could potentially solve the observed discrepancy in the predicted velocity, thus capturing multi-dimensional effects and extending the flow characterization away from the stagnation line.

\section*{Acknowledgments}

The experimental activities of this work were supported by the ESA Contract no. 4000125437/18/NL/RA. The MEEST project (grant N.899298) is acknowledged for funding the experimental campaign for hydrogen Stark broadening measurements. The research of A. Fagnani was funded by the Research Foundation - Flanders (dossier n. 1SB3221N).
The authors would like to acknowledge D.~Luis and Dr.~A.~Viladegut for their collaboration in establishing testing procedures for the hydrogen-seeded mixtures. 

\appendix

\section{Numerical verification of the quasi-1D stagnation line simulation against CF-ICP}
\label{sec:REB_SLFW_validation}

\begin{figure}[h!]
	\centering
	\subfigure[]
	{\includegraphics[width=0.315\linewidth, trim = {0cm, 4.5cm, 0cm, 0cm}]{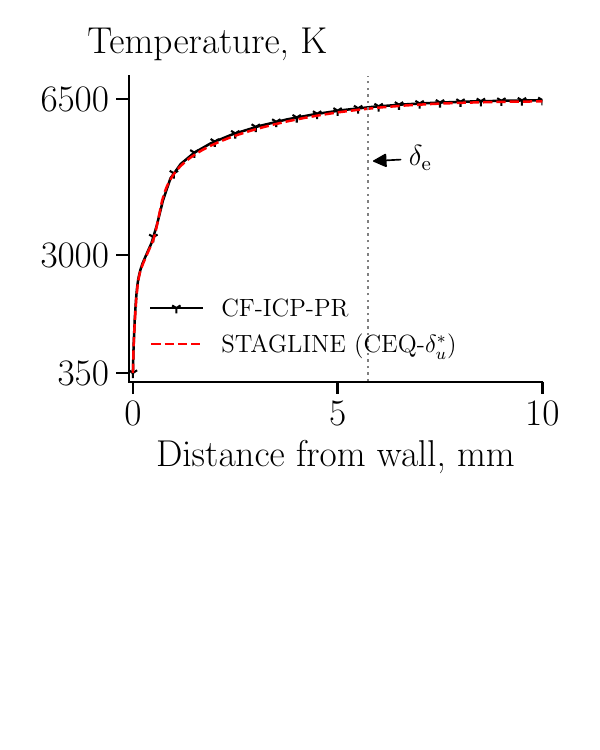} \label{fig:figure4_8a}}
	\subfigure[]
	{\includegraphics[width=0.315\linewidth,trim = {0cm, 4.5cm, 0cm, 0cm}]{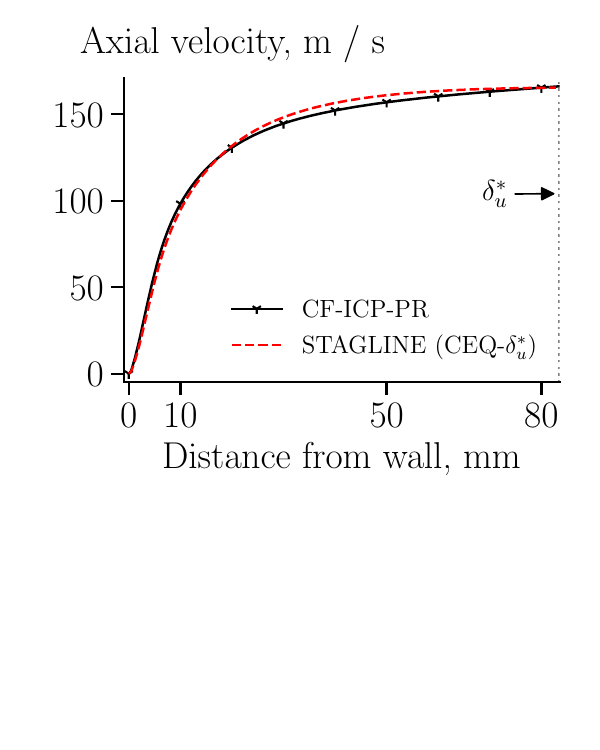}\label{fig:figure4_8b}}
	\subfigure[]
	{\includegraphics[width=0.315\linewidth,trim = {0cm, 4.5cm, 0cm, 0cm}]{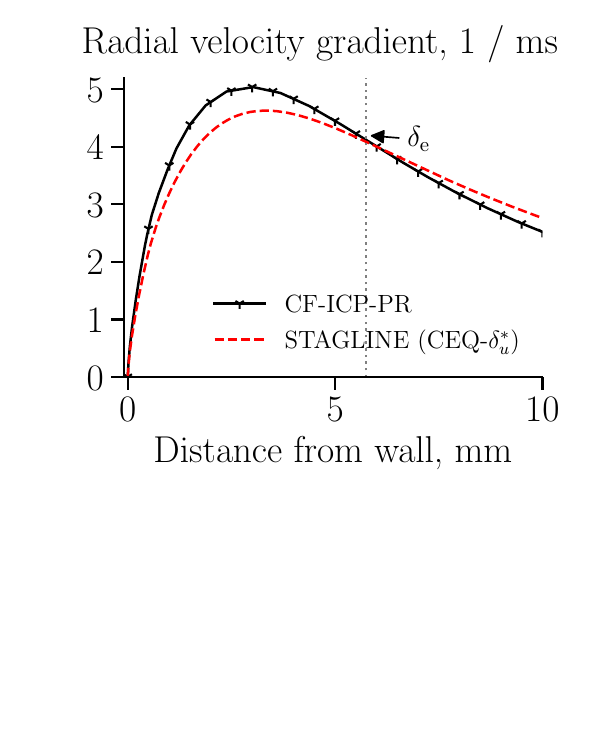}\label{fig:figure4_8c}}\\
	\subfigure[]
	{\includegraphics[trim = {0cm, 0cm, 0cm, 0cm}, clip, width=0.50\linewidth]{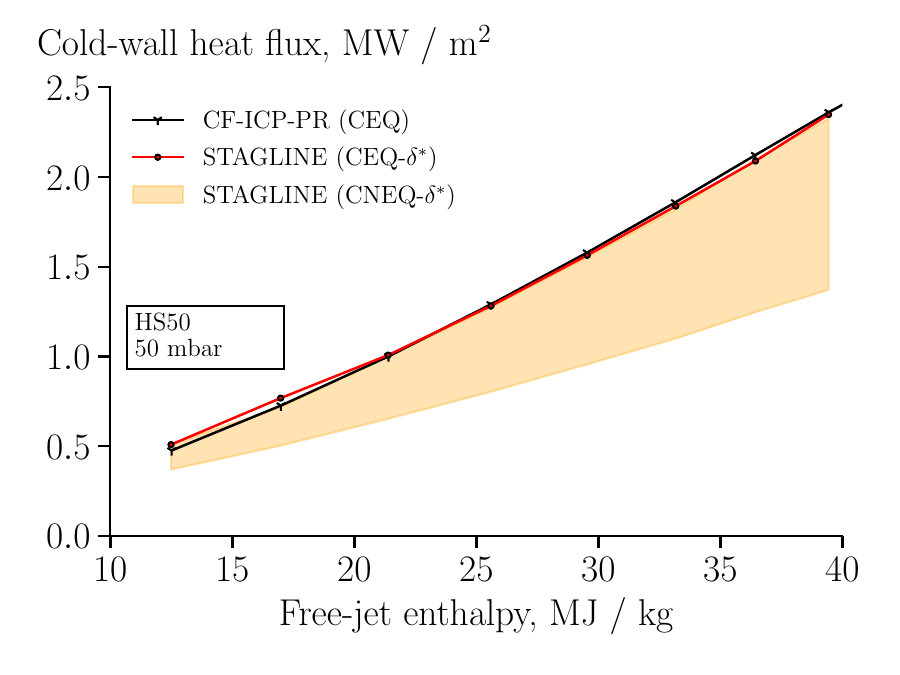}\label{fig:figure4_8d}}
	\subfigure[]
	{\includegraphics[trim = {0cm, 0cm, 0cm, 0cm}, clip, 
		width = 0.40\textwidth]{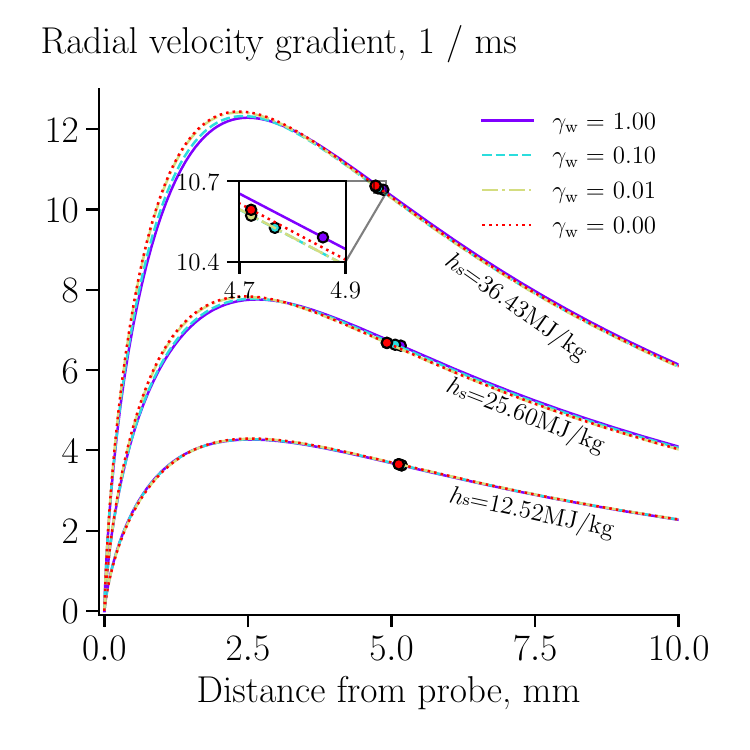}\label{fig:figure4_10b}}
	\caption[]{(a) Temperature, (b) axial velocity, and (c) radial velocity gradient demonstrate limited approximation error againt CF-ICP-PR when the flow is computed with STAGLINE from $ \deltastar $ (50~mbar, 100~kW). (d) CEQ and CNEQ ($ 0<\gamma\subrm{w}<1 $) wall heat flux are compatible with the equilibrium solution of CF-ICP. (e) Radial velocity gradients for different free-jet enthalpy conditions at 50 mbar show negligible dependence on $ \gamma\subrm{w} $. Markers indicate boundary layer edge location at the inflexion point.}
	\label{fig:pc50mbaricpslnebhsqcw}
\end{figure}

A numerical verification was carried out against CF-ICP, in order to evaluate the degree of approximation in the choice of the inlet location and boundary conditions for STAGLINE proposed in sec~\ref{sec:delta_star}. In this case, $ \us $ and $ \Ts $ were extracted from CF-ICP-FJ, and $ \deltastar_u(\us) $ was computed from the correlation given in Fig.~\ref{fig:figure4_7d}, then the 1D stagnation line flow was compared to the 2D CF-ICP-PR solution. 
Figures~\ref{fig:figure4_8a},~\ref{fig:figure4_8b}~and~\ref{fig:figure4_8c} compare the results computed for a condition at $ \pc = \SI{50}{\mbar} $ and $ \Pelsim = \SI{90}{\kW} $.
The temperature predicted by STAGLINE agrees with CF-ICP within $ \pm1\% $ inside the BL, while the approximation on the velocity gradient and axial velocity is somehow poorer, but still limited within $ \pm10\% $. 
We conclude that the thermal boundary layer is closely represented, while the approximation of the momentum is less accurate and likely related to multidimensional effects not represented in STAGLINE.
Fig.~\ref{fig:figure4_8d} further compares the computed cold-wall heat flux, showing that the equilibrium values are consistent, and that the proposed procedure based on STAGLINE provides a good approximation of value predicted by the multidimentional CF-ICP solution.

Additionally, we observe that $ \qcw(\textrm{CEQ})~\approxeq~\qcw(\textrm{CNEQ}, \gammaw = 1) $, i.e., that the CEQ heat flux is an upper limit of the CNEQ value at these conditions.
This supports the point claimed in sec.~\ref{sec:h_q_p_maps}, indicating that CF-ICP was incompatible with intrusive measurements of heat flux due to the underprediction of the axial velocity.

Since the current effort tries to establish a free-jet characterization procedure that is independent of $ \gammaw $, we needed to ensure that $ \betaedge $, which drives the similarity of the boundary layer flow within the LHTS, is also negligibly affected by  $ \gammaw $.
Fig.~\ref{fig:figure4_10b} reports $ \beta $ as a function of the distance to the wall for some sample conditions at low, mid and high values of $ \hs $, confirming that the effect of $ \gammaw $ is negligible.
Thus, knowledge of the chamber pressure, free-jet temperature and velocity, together with the probe geometry, are sufficient for a characterization of the free-jet flow as required by the LHTS.

In summary, the proposed quasi-1D STAGLINE procedure provides a satisfactory approximation of the wall heat flux and boundary layer edge velocity gradient compared to the 2D axisymmetric CF-ICP, which gives confidence for its application to the experimental conditions.

\clearpage
\nolinenumbers
\section{Experimental data for the enthalpy characterization}

\begin{table}[h]
	\small
	\centering
	\caption{Experimental data for $ \pc = \SI{50}{\mbar} $.}
	\label{tab:APP_REB_exp_data_50mbar}
	\resizebox{.5\textwidth}{!}{
		\begin{tabular}{lllll}
			\toprule
			  $ \Pel $ & $ \Ts $ &  $ \qcw $ &  $ \pdyn $ &  $ \qcw $\\
		      $ \SI{}{\kW} $  & $ \SI{}{\kelvin} $ & $ \SI{}{\kW/\meter\squared} $ &  $ \SI{}{\pascal} $ & $ \SI{}{\kW/\meter\squared} $\\
			\midrule
			  &   & HS50 & HS50   &  HS30   \\
			\midrule
			  111 &  N/A &  466 &   20 &       469 \\
			  125 & 5184 &  575 &   26 &       603 \\
			  151 & 5498 &  750 &   42 &       855 \\
			  170 & 5777 & 1154 &   46 &      1214 \\
			  200 & 6116 & 1412 &   60 &      1565 \\
			  252 & 6716 & 2044 &   84 &      2236 \\
			  301 & 7554 & 2577 &  112 &       N/A \\
			  111 & 4848 &  389 &   18 &       495 \\
			  125 & 4996 &  390 &   27 &       625 \\
			  152 & 5426 &  743 &   39 &       935 \\
			  171 & 5710 & 1067 &   44 &      1554 \\
			  200 & 6121 & 1393 &   58 &      1944 \\
			  252 & 6713 & 2033 &   83 &      2681 \\
			  301 & 7513 & 2609 &  110 &      3406 \\
			  126 & 5109 &  560 &   32 &       680 \\
			  151 & 5505 &  823 &   41 &      1173 \\
			  171 &  N/A & 1187 &   47 &      1607 \\
			  202 & 6105 & 1572 &   61 &      2146 \\
		      253 & 6660 & 1873 &   88 &      2569 \\
		      302 & 7512 & 2545 &  113 &      3403 \\
			\bottomrule
		\end{tabular}
	}
\end{table}

\begin{table}[h]
	\small
	\centering
	\caption{Experimental data for $ \pc = \SI{100}{\mbar} $.}
	\label{tab:APP_REB_exp_data_100mbar}
	\resizebox{.45\textwidth}{!}{
		\begin{tabular}{lllll}
			\toprule
			   $ \Pel $ & $ \Ts $ &  $ \qcw $ &  $ \pdyn $ &  $ \qcw $\\
			    $ \SI{}{\kW} $  & $ \SI{}{\kelvin} $ & $ \SI{}{\kW/\meter\squared} $ &  $ \SI{}{\pascal} $ & $ \SI{}{\kW/\meter\squared} $\\
			\midrule
			   &   & HS50 & HS50   &  HS30   \\
			\midrule
			  148 & 5443 &  740 &   15 &       877 \\
			  172 & 5417 &  722 &   21 &       930 \\
			  190 & 5733 &  854 &   23 &      1061 \\
			  211 & 5923 & 1307 &   27 &      1524 \\
		 	  226 & 6091 & 1594 &   29 &      1880 \\
			  252 & 6325 & 2029 &   39 &      2370 \\
			  269 & 6524 & 2222 &   44 &      2546 \\
			  290 & 6779 & 2470 &   55 &      2730 \\
			  190 &  N/A & 1273 &   24 &      1503 \\
			  153 & 5422 &  741 &   16 &       918 \\
			  172 &  N/A &  861 &   17 &      1025 \\
			  172 &  N/A &  766 &   17 &         0 \\
			 256 &  N/A & 2026 &   44 &      2314 \\
			  152 & 5415 &  753 &   16 &       877 \\
			  173 & 5646 &  814 &   17 &       966 \\
			  192 & 5830 & 1027 &   22 &      1281 \\
			  209 & 6174 & 1398 &   27 &      1639 \\
			  253 & 6359 & 1982 &   42 &      2334 \\
			  289 & 6752 & 2518 &   56 &      3053 \\
			  173 & 5623 &  777 &   21 &       902 \\
			  190 & 5899 & 1030 &   21 &      1168 \\
			  209 & 6134 & 1407 &   26 &      1807 \\
			  253 & 6365 & 1992 &   42 &      2368 \\
			  292 & 6889 & 2744 &   60 &      3276 \\
		 	 189 & 6026 & 1266 &   25 &      1441 \\
			 171 & 5753 &  874 &   20 &      1010 \\
			  173 & 5656 &  834 &   18 &      1117 \\
			  191 & 5937 & 1143 &   24 &      1554 \\
			  211 & 6100 & 1589 &   30 &      2079 \\
			  150 & 5514 &  825 &   16 &      1156 \\
			  174 & 5622 &  964 &   20 &      1211 \\
			  191 & 5984 & 1241 &   26 &      1737 \\
			  210 & 6144 & 1660 &   33 &      2274 \\
			 233 & 6312 & 2027 &   38 &      2630 \\
			 253 &  N/A & 2182 &   44 &      2924 \\
		 	 249 & 6453 & 2181 &   44 &      3017 \\
			  269 & 6641 & 2318 &   51 &      3264 \\
			  292 & 6819 & 2683 &   62 &      3651 \\
			\bottomrule
		\end{tabular}
	}
\end{table}

\section{Electronic transitions}

\begin{table}[h!]
	\centering
	\caption{Atomic transitions considered in this work. Data from the NIST Atomic Spectra Database \cite{NISTASD}.}
	\resizebox{0.7\textwidth}{!}{ 
		\small
		\begin{tabular}{llllllllll}
			\hline
			ID & Wavelength & Upper & Lower & $ g_l $ & $ g_u $ & $ \einsteincoeff $ & $ E_l $ & $ E_u $ & Acc. \\
			& $\mathrm{\mathring{A}}$ & level  & level  &  &  & $\mathrm{s^{-1}}$ & $\mathrm{eV}$ & $\mathrm{eV}$ & \% \\
			\hline
			\multirow{7}{*}{H486} & 4861.279 & $ \specnot{4d}{2}{D}{3/2}{} $ & $ \specnot{2p}{2}{P}{1/2}{*} $ & 2 & 4 & 1.7188e+07 & 10.199 & 12.749 & 0.3 \\
			& 4861.287 & $ \specnot{4p}{2}{P}{3/2}{*} $ & $ \specnot{2s}{2}{S}{1/2}{} $ & 2 & 4 & 9.6680e+06 & 10.199 & 12.749 & 0.3 \\
			& 4861.288 & $ \specnot{4s}{2}{S}{1/2}{} $ & $ \specnot{2p}{2}{P}{1/2}{*} $ & 2 & 2 & 8.5941e+05 & 10.199 & 12.749 & 0.3 \\
			& 4861.298 & $ \specnot{4p}{2}{P}{1/2}{*} $ & $ \specnot{2s}{2}{S}{1/2}{} $ & 2 & 2 & 9.6682e+06 & 10.199 & 12.749 & 0.3 \\
			& 4861.362 & $ \specnot{4d}{2}{D}{5/2}{} $ & $ \specnot{2p}{2}{P}{3/2}{*} $ & 4 & 6 & 2.0625e+07 & 10.199 & 12.749 & 0.3 \\
			& 4861.365 & $ \specnot{4d}{2}{D}{3/2}{} $ & $ \specnot{2p}{2}{P}{3/2}{*} $ & 4 & 4 & 3.4375e+06 & 10.199 & 12.749 & 0.3 \\
			& 4861.375 & $ \specnot{4s}{2}{S}{1/2}{} $ & $ \specnot{2p}{2}{P}{3/2}{*} $ & 4 & 2 & 1.7190e+06 & 10.199 & 12.749 & 0.3 \\
			\hline
			\multirow{7}{*}{H656} & 6562.710 & $ \specnot{3d}{2}{D}{3/2}{} $ & $ \specnot{2p}{2}{P}{1/2}{*} $ & 2 & 4 & 5.3877e+07 & 10.199 & 12.088 & 0.3 \\
			& 6562.725 & $ \specnot{3p}{2}{P}{3/2}{*} $ & $ \specnot{2s}{2}{S}{1/2}{} $ & 2 & 4 & 2.2448e+07 & 10.199 & 12.088 & 0.3 \\
			& 6562.752 & $ \specnot{3s}{2}{S}{1/2}{} $ & $ \specnot{2p}{2}{P}{1/2}{*} $ & 2 & 2 & 2.1046e+06 & 10.199 & 12.087 & 0.3 \\
			& 6562.772 & $ \specnot{3p}{2}{P}{1/2}{*} $ & $ \specnot{2s}{2}{S}{1/2}{} $ & 2 & 2 & 2.2449e+07 & 10.199 & 12.087 & 0.3 \\
			& 6562.852 & $ \specnot{3d}{2}{D}{5/2}{} $ & $ \specnot{2p}{2}{P}{3/2}{*} $ & 4 & 6 & 6.4651e+07 & 10.199 & 12.088 & 0.3 \\
			& 6562.867 & $ \specnot{3d}{2}{D}{3/2}{} $ & $ \specnot{2p}{2}{P}{3/2}{*} $ & 4 & 4 & 1.0775e+07 & 10.199 & 12.088 & 0.3 \\
			& 6562.909 & $ \specnot{3s}{2}{S}{1/2}{} $ & $ \specnot{2p}{2}{P}{3/2}{*} $ & 4 & 2 & 4.2097e+06 & 10.199 & 12.087 & 0.3 \\
			\hline
			\multirow{9}{*}{O532} & 5329.096 & $ \specnot{5d}{5}{D}{0}{*} $ & $ \specnot{3p}{5}{P}{1}{} $ & 3 & 1 & 2.7100e+06 & 10.740 & 13.066 & 18.0 \\
			& 5329.099 & $ \specnot{5d}{5}{D}{1}{*} $ & $ \specnot{3p}{5}{P}{1}{} $ & 3 & 3 & 2.0300e+06 & 10.740 & 13.066 & 18.0 \\
			& 5329.107 & $ \specnot{5d}{5}{D}{2}{*} $ & $ \specnot{3p}{5}{P}{1}{} $ & 3 & 5 & 9.4800e+05 & 10.740 & 13.066 & 18.0 \\
			& 5329.673 & $ \specnot{5d}{5}{D}{1}{*} $ & $ \specnot{3p}{5}{P}{2}{} $ & 5 & 3 & 6.7700e+05 & 10.740 & 13.066 & 18.0 \\
			& 5329.681 & $ \specnot{5d}{5}{D}{2}{*} $ & $ \specnot{3p}{5}{P}{2}{} $ & 5 & 5 & 1.5800e+06 & 10.740 & 13.066 & 18.0 \\
			& 5329.690 & $ \specnot{5d}{5}{D}{3}{*} $ & $ \specnot{3p}{5}{P}{2}{} $ & 5 & 7 & 1.8100e+06 & 10.740 & 13.066 & 18.0 \\
			& 5330.726 & $ \specnot{5d}{5}{D}{2}{*} $ & $ \specnot{3p}{5}{P}{3}{} $ & 7 & 5 & 1.8000e+05 & 10.741 & 13.066 & 18.0 \\
			& 5330.735 & $ \specnot{5d}{5}{D}{3}{*} $ & $ \specnot{3p}{5}{P}{3}{} $ & 7 & 7 & 9.0200e+05 & 10.741 & 13.066 & 18.0 \\
			& 5330.741 & $ \specnot{5d}{5}{D}{4}{*} $ & $ \specnot{3p}{5}{P}{3}{} $ & 7 & 9 & 2.7100e+06 & 10.741 & 13.066 & 18.0 \\
			\hline
			\multirow{9}{*}{O615} & 6155.961 & $ \specnot{4d}{5}{D}{0}{*} $ & $ \specnot{3p}{5}{P}{1}{} $ & 3 & 1 & 7.6200e+06 & 10.740 & 12.754 & 7.0 \\
			& 6155.971 & $ \specnot{4d}{5}{D}{1}{*} $ & $ \specnot{3p}{5}{P}{1}{} $ & 3 & 3 & 5.7200e+06 & 10.740 & 12.754 & 7.0 \\
			& 6155.989 & $ \specnot{4d}{5}{D}{2}{*} $ & $ \specnot{3p}{5}{P}{1}{} $ & 3 & 5 & 2.6700e+06 & 10.740 & 12.754 & 7.0 \\
			& 6156.737 & $ \specnot{4d}{5}{D}{1}{*} $ & $ \specnot{3p}{5}{P}{2}{} $ & 5 & 3 & 1.9100e+06 & 10.740 & 12.754 & 7.0 \\
			& 6156.755 & $ \specnot{4d}{5}{D}{2}{*} $ & $ \specnot{3p}{5}{P}{2}{} $ & 5 & 5 & 4.4500e+06 & 10.740 & 12.754 & 7.0 \\
			& 6156.778 & $ \specnot{4d}{5}{D}{3}{*} $ & $ \specnot{3p}{5}{P}{2}{} $ & 5 & 7 & 5.0800e+06 & 10.740 & 12.754 & 7.0 \\
			& 6158.149 & $ \specnot{4d}{5}{D}{2}{*} $ & $ \specnot{3p}{5}{P}{3}{} $ & 7 & 5 & 5.0700e+05 & 10.741 & 12.754 & 7.0 \\
			& 6158.172 & $ \specnot{4d}{5}{D}{3}{*} $ & $ \specnot{3p}{5}{P}{3}{} $ & 7 & 7 & 2.5400e+06 & 10.741 & 12.754 & 7.0 \\
			& 6158.187 & $ \specnot{4d}{5}{D}{4}{*} $ & $ \specnot{3p}{5}{P}{3}{} $ & 7 & 9 & 7.6200e+06 & 10.741 & 12.754 & 7.0 \\
			\hline
			\multirow{3}{*}{O777} & 7771.944 & $ \specnot{3p}{5}{P}{3}{} $ & $ \specnot{3s}{5}{S}{2}{*} $ & 5 & 7 & 3.6900e+07 & 9.146 & 10.741 & 3.0 \\
			& 7774.166 & $ \specnot{3p}{5}{P}{2}{} $ & $ \specnot{3s}{5}{S}{2}{*} $ & 5 & 5 & 3.6900e+07 & 9.146 & 10.740 & 3.0 \\
			& 7775.388 & $ \specnot{3p}{5}{P}{1}{} $ & $ \specnot{3s}{5}{S}{2}{*} $ & 5 & 3 & 3.6900e+07 & 9.146 & 10.740 & 3.0 \\
			\hline
			\multirow{3}{*}{O844} & 8446.247 & $ \specnot{3p}{3}{P}{0}{} $ & $ \specnot{3s}{3}{S}{1}{*} $ & 3 & 1 & 3.2200e+07 & 9.521 & 10.989 & 10.0 \\
			& 8446.359 & $ \specnot{3p}{3}{P}{2}{} $ & $ \specnot{3s}{3}{S}{1}{*} $ & 3 & 5 & 3.2200e+07 & 9.521 & 10.989 & 10.0 \\
			& 8446.758 & $ \specnot{3p}{3}{P}{1}{} $ & $ \specnot{3s}{3}{S}{1}{*} $ & 3 & 3 & 3.2200e+07 & 9.521 & 10.989 & 10.0 \\
			\hline
			\multirow{1}{*}{N746} & 7468.312 & $ \specnot{3p}{4}{S}{3/2}{*} $ & $ \specnot{3s}{4}{P}{5/2}{} $ & 6 & 4 & 1.9600e+07 & 10.336 & 11.996 & 7.0 \\
			\hline
			\multirow{3}{*}{N868} & 8680.282 & $ \specnot{3p}{4}{D}{7/2}{*} $ & $ \specnot{3s}{4}{P}{5/2}{} $ & 6 & 8 & 2.5300e+07 & 10.336 & 11.764 & 7.0 \\
			& 8683.403 & $ \specnot{3p}{4}{D}{5/2}{*} $ & $ \specnot{3s}{4}{P}{3/2}{} $ & 4 & 6 & 1.8800e+07 & 10.330 & 11.758 & 7.0 \\
			& 8686.149 & $ \specnot{3p}{4}{D}{3/2}{*} $ & $ \specnot{3s}{4}{P}{1/2}{} $ & 2 & 4 & 1.1500e+07 & 10.326 & 11.753 & 7.0 \\
			\hline
		\end{tabular}
	}
	\label{tab:APP_OES_atomic_transitions}
\end{table}

\clearpage

\bibliographystyle{elsarticle-num-names} 
\bibliography{biblio}

\end{document}